\documentclass[a4paper,fleqn,usenatbib]{mnras}
\usepackage[T1]{fontenc}
\usepackage{ae,aecompl}
\usepackage{comment}
\usepackage{soul}
\usepackage[dvipsnames]{xcolor}
\usepackage{graphicx}	% Including figure files
\usepackage{amsmath}	% Advanced maths commands
\usepackage{amssymb}	% Extra maths symbols
\usepackage{mathrsfs}
\usepackage{listings}
\usepackage{cuted}
\usepackage{flushend}
\usepackage{todonotes}
\usepackage{lineno}
\usepackage{enumitem,cleveref}
\usepackage{fancybox,framed}
\usepackage{newtxtext,newtxmath}
\makeatletter
\preto\gather{\@fleqnfalse}
\makeatother
\DeclareMathOperator{\sinc}{sinc}
\DeclareFontFamily{U}{wncy}{}
\DeclareFontShape{U}{wncy}{m}{n}{<->wncyr10}{}
\DeclareSymbolFont{mcy}{U}{wncy}{m}{n}
\DeclareMathSymbol{\Sh}{\mathord}{mcy}{"58}

\title[High accuracy wide field imaging method]{High accuracy wide field imaging method in radio interferometry}

\author[H. Ye et al.]{
Haoyang Ye,$^{1}$\thanks{E-mail: haoyang.ye@cantab.net}
Stephen F. Gull,$^{2}$\thanks{E-mail: sfg1@cam.ac.uk}
Sze M. Tan $^{3}$
Bojan Nikolic$^{2}$
\\
$^{1}$School of Astronomy and Space Science, Nanjing University, Nanjing 210093, China\\
$^{2}$Astrophysics Group, Cavendish Laboratory, University of Cambridge, Cambridge CB3 0HE, UK\\
$^{3}$Picarro, Inc., 3105 Patrick Henry Dr., Santa Clara, CA 95054, USA
}

\date{Submitted to MNRAS. Accepted XXX. Received YYY; in original form ZZZ}

\pubyear{2021}
\begin{document}
\label{firstpage}
\pagerange{\pageref{firstpage}--\pageref{lastpage}}
\maketitle

\begin{abstract}
With the development of modern radio interferometers, wide-field continuum surveys have been planned and undertaken, for which accurate wide-field imaging methods are essential.
Based on the widely-used W-stacking method, we propose a new wide-field imaging algorithm that can synthesize visibility data from a model of the sky brightness via degridding,
able to construct dirty maps from measured visibility data via gridding. Results carry the smallest approximation error yet achieved relative to the exact calculation involving the direct Fourier transform. In contrast to the original W-stacking method, the new algorithm performs least-misfit optimal gridding (and degridding) in all three directions, and is capable of achieving much higher accuracy than is feasible with the original algorithm. In particular, accuracy at the level of single precision arithmetic is readily achieved by choosing a least-misfit convolution function of width $W=7$ and an image cropping parameter of $x_0=0.25$. If the accuracy required is only that attained by the original W-stacking method, the computational cost for both the gridding and FFT steps can be substantially reduced using the proposed method by making an appropriate choice of the width and image cropping parameters.
\end{abstract}

\begin{keywords}
techniques: interferometric -- techniques: image processing -- methods: analytical -- methods: observational -- methods: data analysis
\end{keywords}
%%%%%%%%%%%%%%%%%%%%%%%%%%%%%%%%%%%%%%%%%%%%%%%%%%

%%%%%%%%%%%%%%%%% BODY OF PAPER %%%%%%%%%%%%%%%%%%

\section{Introduction}

With the continuing development of radio telescopes
and data processing techniques, radio observations have facilitated cutting-edge scientific discoveries in astronomy. They also provide information available only at radio frequencies that gives insight into a wide range of astrophysical phenomena.

A radio interferometer
%is a radio telescope that
combines multiple antennas electronically in collecting radio emissions from the same direction of the sky. The %geographic
separation of each pair of antennas constitutes a `baseline'. The longest baseline is between the two antennas that are farthest apart; its length is inversely proportional to the angular resolution of the observation, making radio interferometers ideal for high-resolution radio observations.

Several next-generation or upgraded radio interferometers
are currently undertaking or planning wide-field continuum surveys all over the world.
These include EMU \citep{2020AAS...23632206K} by ASKAP (Australian Square
Kilometre Array Pathfinder), LoTSS \citep{2019A&A...622A...1S} by
LoFAR (Low--Frequency Array), as well as VLASS
\citep{2020PASP..132c5001L} by the upgraded VLA and the Apertif radio
continuum survey \citep{2020AAS...23513607A} by WSRT with the new
phased-array feed APERTIF (APERture Tile In Focus). %The large area covered by each pointing of these surveys, such as 30 square degrees for ASKAP and 67 square degrees for LoFAR,
The success of these wide-field continuum surveys requires accurate and efficient wide-field imaging methods. One of the challenges for wide-field imaging is that the diffraction of the wavefront between elements of the interferometer is significantly different in different observed directions. This is termed the `$w$-effect' or `non-coplanar effect'. Algorithms such as the W-projection method \citep{Cornwell2003W, 2005ASPC..347...86C}, the W-snapshot method \citep{2012SPIE.8500E..0LC} and the W-stacking method \citep{humphreys2011analysis, 2014MNRAS.444..606O} have been proposed. The W-projection method is implemented in a widely used imaging pipeline \texttt{CASA} \citep{2007ASPC..376..127M}), and has been widely applied in the imaging process; see, for example, \citet{2016ApJ...833...12R, 2017NatAs...1E...5V}. The W-snapshot and W-stacking methods are implemented in \texttt{WSCLEAN}. The W-stacking method has been used by \citet{2016ApJ...827L..22T, 2018MNRAS.478.2218H}. \citet{2019ApJ...874..174P} later proposed a hybrid W-stacking and W-projection algorithm. Efforts have also been made to perform wide-field imaging using compressed sensing techniques \citep{2011MNRAS.413.1318M}.

The present paper proposes a new wide-field imaging algorithm which gives the smallest wide-field degridding and dirty image-making approximation error yet achieved (i.e., except for the direct Fourier transform) at a practical computational cost. This algorithm would provide great benefit in the upcoming radio continuum surveys, especially where high accuracy is needed.
%It would be of great benefit to the upcoming radio continuum surveys, as well as the surveys' sub-fields imaging and other wide-field observations where great imaging accuracy is highly demanded.

This algorithm is based on the widely-used W-stacking method \citep{humphreys2011analysis, 2014MNRAS.444..606O}, but utilizes a three-dimensional gridding process which we shall demonstrate can attain considerably higher accuracy than the W-stacking method as currently practised. In this new algorithm, the number of `$w$- layers' involved can also be greatly reduced, greatly reducing the number of operations of the Fast Fourier Transform (FFT, \citet{cooley1965algorithm}). To achieve an accuracy of $10^{-7}$, or single precision accuracy, we recommend using the `least-misfit gridding function' with a width of $W=7$ and an image cropping parameter of $x_0=0.25$. However, if an accuracy close to that of the original W-stacking method is required, the use of three-dimensional gridding with appropriately chosen values of $W$ and $x_0$ can significantly reduce the computational cost for both the gridding step and the FFT step.

Our algorithm has been implemented in the \texttt{WSCLEAN} software \citep{2014MNRAS.444..606O} as a module called `WGridder' and \texttt{NIFTY} \citep{2013A&A...554A..26S,
  2017arXiv170801073S} by \citet{2020arXiv201010122A}. A \texttt{Python}-based tutorial is also available \footnote{\url{https://github.com/zoeye859/Wide-field-Nfacet}}.

For wide-field radio observation, it is vital also to remove the direction dependent effects (DDEs) due to observation instruments (such as pointing errors, primary beam variations) or by the ionosphere, which often corrupt the observation. Imaging methods such as the A-Projection algorithm \citep{2013ApJ...770...91B} and the facet-based SSD (SubSpace Deconvolution) algorithm \citep{2018A&A...611A..87T} have been devised to deal with DDEs. The present algorithm can be combined with the A-Projection algorithm for DDE corrections.

In Section \ref{s:fundamental}, below, we set out the essentials of the wide-field imaging problem. Section \ref{sec:error_bounds} provides a theoretical framework for evaluating and comparing the accuracy of methods for calculating expected visibility data from models of sky brightness, and for making maps from visibility measurements. Section \ref{s:imaging} explains the W-stacking algorithm, and Section \ref{sec:W-stacking} describes our proposed `improved W-stacking method' in full detail. Section \ref{sec:performance} quantifies the performance of our `Improved W-stacking method' by showing the approximation errors and computational cost. We then demonstrate various applications using both simulated and real data in Section \ref{sec:application}, and present conclusions in Section \ref{sec:conclusion}.

\section{Fundamentals of wide--field imaging}\label{s:fundamental}

In a radio interferometer, the electric field is measured on a set of antennas. The vector separating a pair of antennas is termed a baseline $(u,v,w)$, where $u$, $v$ and $w$ are measured in wavelengths and $w$ is directed along the line of sight to the phase center of the observation. The (complex) cross-correlation of the electric fields received at these antennas is called the `visibility' on this baseline, and is denoted $V(u,v,w)$. The sky brightness distribution function $I(l,m)$ represents the radio emission strength of the incoherent sources on the celestial sphere, where $l$, $m$ and $n=\sqrt{1-l^2-m^2}$ are direction cosines to the far-field source from the point of observation. The collected visibility data, $V(u,v,w)$, can be written as a function of $I(l,m)$ as:
\begin{linenomath}
\begin{multline}\label{eq:fourier_visibility}
	V(u,v,w) = \iint \text{d}l \text{d}m \,
	\frac{I(l,m)}{n}\exp\left[-i2\pi\left(ul+vm+w\left(n-1\right)\right)\right]
\end{multline}
\end{linenomath}
When the field of view is so small that $n=\sqrt{1-l^2-m^2}\approx 1$ over the entire field, and the baselines are such that the so-called `$w$-term' $w(n-1)\approx 0$ in the exponent can be ignored, this equation reduces to a two-dimensional Fourier transform of visibilities on the $(u,v)$ plane. If it were possible to make complete measurements over the entire $(u,v)$ plane, the sky intensity function $I(l,m)$ could be recovered from the visibilities via a two-dimensional inverse Fourier transform.

As the measurements are incomplete and available only on a discrete set of baselines, a discrete inverse Fourier transform of the measured visibilities results not in the true sky brightness but a `dirty image'. This is equal to the convolution between the true sky
brightness and the Fourier transform of the $uv$-plane sampling
function; the latter is also known as the `synthesised' beam or `dirty
beam'. A deconvolution algorithm such as CLEAN
\citep{1974A&AS...15..417H,1980A&A....89..377C,1983AJ.....88..688S,2008ISTSP...2..793C}
or the Maximum Entropy method (MEM) \citep{1978Natur.272..686G,
  gull1984maximum, gull1989developments} is needed to deconvolve the
true sky brightness from the dirty image by means of an iterative
process.

As the field of view increases, or if the interferometer geometry means it is invalid to approximate the portion of the celestial sphere imaged as a plane parallel to a plane formed by the baselines, the $w$-terms cannot be neglected. The imaging process described above then fails to reconstruct the true sky brightness, and imaging methods capable of dealing with these extra $w$-terms are required. Such methods are known as wide field imaging methods.

The Fresnel number $N_F=\frac{D^2}{B\lambda}$ is useful in determining whether the $w$-term must be considered \citep{2005ASPC..347...86C}. Here, $D$ is the effective diameter of each antenna, so that the angular field of view is $\approx\lambda/D$, and $B$ is the maximum extent of any baseline in the $w$ direction, while $\lambda$ is the observing wavelength. When $N_F<1$, a wide-field imaging method is needed in order to relate sky brightness and visibilities accurately. Therefore, for arrays whose antennas are not co-planar having large $w$ values (and correspondingly a large value  of $B$), and especially for low frequency observations for which $\lambda$ is large, wide field imaging methods are required.

\section{Image Reconstruction and Mapping; the roles of Degridding and Gridding}\label{sec:error_bounds}

In radio interferometry data processing, the data consist of measurements of visibility data ${V}_k$ on a finite number of baselines. The image reconstruction problem involves generating a model of the sky brightness function $I(l,m)$ which is {\em feasible} in the sense that it fits the measured data to within the measurement uncertainties. As a result of the incompleteness of the measurements, there are often infinitely many feasible models, and some criterion of optimality such as the maximum entropy method is required to select from among these. A primary objective of this paper is to demonstrate how to calculate the data misfit accurately in the context of wide-field imaging, since this is central to all reconstruction algorithms which monitor feasibility as they converge towards an optimal solution.

\subsection{The adjoint relation between calculating visibilities from a model sky and aperture synthesis mapping}

In this section we show that the operations of calculating visibilities from a model of sky brightness and of constructing a dirty image from incomplete visibility data have an adjoint relationship. These operations are both computationally expensive when the baselines are not on a uniform rectangular grid. That is because they involve discrete Fourier transforms (DFT), but they can be approximated by more efficient methods which use the fast Fourier transform (FFT) algorithm, together with operations known as degridding and gridding respectively. It will emerge that it is highly advantageous to maintain the adjoint relationship when making these approximations, since this allows us to obtain common bounds for both approximation errors.

We shall be using notation and concepts from the theory of Hilbert spaces in functional analysis, which are described in detail in references such as \citet{Conway1990_FunctionalAnalysis} and summarized below.

Let us first work in one dimension, postponing the full three dimensional case to a subsequent section. Assuming that the sky brightness $I(l)$ is restricted to $[-l_{\rm{max}}, l_{\rm{max}}]$, the visibility ${V}_k$ on a baseline $u_k$ is
\begin{linenomath}
\begin{equation}
{V}_k = \int_{-l_{\rm{max}}}^{l_{\rm{max}}}
I(l)\exp(-i2\pi u_kl)\,\mathrm{d}l
\label{eq:fwd_1d_unnorm}
\end{equation}
\end{linenomath}
In practice, we work with brightness distributions (`images') on a regular grid and the integral is implemented as a finite sum, although this distinction is not important in the following analysis. The gridding and degridding processes of interest take place in visibility space, and it is convenient to continue treating image space as consisting of functions of a continuous variable. We shall also use the normalized map coordinate $x$, which is related to the direction cosine $l$ by $x/x_0 = l/l_{\mathrm{max}}$, where $x_0\leq\frac{1}{2}$. For reasons that will become apparent in the description of Algorithm 1 below, we refer to $x_0$ as the image cropping parameter.

The variable conjugate to $x$ is $u'=l_{\mathrm{max}}u/x_0$. Whereas $u$ is measured in wavelengths, $u'$ is such that integer values of $u'$ will correspond to the gridpoints in visibility space associated with the FFT algorithm. Upon changing the variable of integration in (\ref{eq:fwd_1d_unnorm}), we find
\begin{linenomath}
\begin{equation}
{V}_k = \frac{l_{\rm{max}}}{x_0}\int_{-x_0}^{x_0}
f(x)\exp(-i2\pi u'_kx)\,\mathrm{d}x
\label{eq:fwd_1d_norm}
\end{equation}
\end{linenomath}
where $f(x)\equiv I(x l_{\rm{max}}/x_0)$. By denoting the vector of visibilities by $\mathbf{V}$ and the function $x\mapsto f(x)$ by $\mathbf{f}$,
(\ref{eq:fwd_1d_norm}) is written in operator form as
\begin{linenomath}
\begin{equation}
\mathbf{V} = \mathbf{A}\mathbf{f}
\label{eq:fwd_1d_op}
\end{equation}
\end{linenomath}
The operator $\mathbf{A}$ is a continuous linear operator between two Hilbert spaces. The first is the image space of square-integrable functions on $[-x_0,x_0]$ upon which we define the inner product
\begin{linenomath}
\begin{equation}
\langle \mathbf{f}, \mathbf{g} \rangle =
\frac{1}{2x_0}\int_{-x_0}^{x_0}f(x)g^*(x)\,\mathrm{d}x
\label{eq:image1d_ip}
\end{equation}
\end{linenomath}
The second consists of vectors with $M$ components (the number of baselines) in visibility space, with the inner product
\begin{linenomath}
\begin{equation}
\langle \mathbf{U},  \mathbf{V}\rangle =
\frac{1}{M}\sum_{k=1}^M {U}_k {V}^*_k
\label{eq:natural_vis_ip}
\end{equation}
\end{linenomath}
The operator $\mathbf{A}$ can be represented as a generalized matrix having one discrete and one continuous index with elements $A_k(x)$, such that
\begin{linenomath}
\begin{equation}
{V}_k = \int_{-x_0}^{x_0}A_k(x) f(x)\,\mathrm{d}x
\label{eq:fwd_operator_1d}
\end{equation}
\end{linenomath}
where $A_k(x) = l_{\mathrm{max}}\exp(-i2\pi u'_k x)/x_0.$
The norm of an element in a Hilbert space is defined in terms of the inner product on that space by
\begin{equation}
\lVert\mathbf{f}\rVert = \sqrt{\langle \mathbf{f},  \mathbf{f} \rangle}
\end{equation}
The norm of an operator (such as $\mathbf{A}$) between the two Hilbert spaces is defined as
\begin{linenomath}
\begin{equation}
\lVert\mathbf{A}\rVert = \sup_{\mathbf{f}\neq 0}
\frac{\lVert\mathbf{Af}\rVert}{\lVert\mathbf{f}\rVert}
\end{equation}
\end{linenomath}
where the norms on the right hand side are in each of the appropriate Hilbert spaces.

The {\em adjoint} of the operator is denoted $\mathbf{A}^\dagger$, and is defined such that
\begin{linenomath}
\begin{equation}
\langle\mathbf{Af}, \mathbf{V}\rangle =
\langle\mathbf{f}, \mathbf{A}^\dagger\mathbf{V}\rangle
\end{equation}
\end{linenomath}
for all elements $\mathbf{f}$ and $\mathbf{V}$ in the respective spaces.

From the definitions above, it is easy to show that, for the operator $\mathbf{A}$ of 1-d mapping defined above,
\begin{linenomath}
\begin{equation}
(\mathbf{A}^\dagger\mathbf{V})(x) = 2l_\mathrm{max}\left(\frac{1}{M}\sum_{k=1}^M {V}_k \exp(i2\pi u'_k x)\right)
\label{eq:dirty_1d}
\end{equation}
\end{linenomath}
We recognize the right hand side as $2l_{\mathrm{max}} d(x)$, where $d(x)$ is the naturally-weighted dirty image of conventional radio-astronomical processing which is the DFT of the visibilities. The weights applied to the visibilities are $1/M$ and sum to $1$. It is also possible to treat more general weighting schemes by redefining the inner product.

\subsection{Gridding and degridding as operator approximations}

Suppose now that we wish to approximate the operator $\mathbf{A}$ by another continuous linear operator $\widetilde{\mathbf{A}}$. This induces an approximation of $\mathbf{A}^\dagger$ by $\widetilde{\mathbf{A}}^\dagger$.

The relevant result in Hilbert space theory (see for example Chapter 2 of \citet{Conway1990_FunctionalAnalysis}) is the equality of the norms of a continuous linear operator and of its adjoint. Since $\mathbf{A}-\widetilde{\mathbf{A}}$ and
$\mathbf{A}^\dagger - \widetilde{\mathbf{A}}^\dagger$
are adjoints, it follows that
\begin{linenomath}
\begin{equation}
\left\lVert\mathbf{A}-\widetilde{\mathbf{A}}\right\rVert =
\left\lVert\mathbf{A}^\dagger - \widetilde{\mathbf{A}}^\dagger\right\rVert
\label{eq:equal_norms}
\end{equation}
\end{linenomath}

As shown in (\ref{eq:dirty_1d}), finding ${\mathbf{A}}^\dagger\mathbf{V}$ involves a calculation of the dirty map via the DFT. The approximation
$\widetilde{\mathbf{A}}^\dagger$ that we wish to consider is the calculation of the dirty map via convolutional gridding followed by grid correction. As summarized in Algorithm 1, this involves the introduction of an even gridding convolution function $C(u')$ having
compact support (meaning that $C(u')$ is nonzero only within the interval $-W/2\leq u' < W/2$ for some integer $W$ which defines its \emph{width}), and a gridding correction function $h(x)$. The approximation to (\ref{eq:dirty_1d}) is
\begin{linenomath}
\begin{align}
(\widetilde{\mathbf{A}}^\dagger\mathbf{V})(x) &=
2l_\mathrm{max} \widetilde{d}(x) \\
&= 2l_\mathrm{max}h(x)\frac{1}{M}  \sum_{r\in\mathbb{Z}}  \left[\sum_{k=1}^M{V}_k C(r-u'_k)\right] \exp(i2\pi r x)
\label{eq:dirty_1d_approx}
\end{align}
\end{linenomath}
Here, $\widetilde{d}(x)$ is the dirty map calculated using gridding. The term in square brackets represents the convolutional gridding, and the sum over $r$ represents the computation of the FFT. The sum over $r$ is finite because of the compactness of the support of $C$ and the maximum baseline length.

The norm of the difference between the operators is
\begin{linenomath}
\begin{align}
\left\lVert\mathbf{A}^\dagger-\widetilde{\mathbf{A}}^\dagger\right\rVert &= 2l_\mathrm{max}\sup_{\mathbf{V}\neq 0}
\frac{\lVert d(x)-\widetilde{d}(x)\rVert}{\lVert\mathbf{V}\rVert}\\
&=2l_\mathrm{max}\sup_{\mathbf{V}\neq 0}
\frac{\displaystyle\left(\frac{1}{2x_0}\int_{-x_0}^{x_0} |d(x)-\widetilde{d}(x)|^2\,\mathrm{d}x\right)^\frac{1}{2}}
{\displaystyle\left(\frac{1}{M}\sum_{k=1}^M|{V}_k|^2\right)^\frac{1}{2}}
\label{eq:norm_diff_1}
\end{align}
\end{linenomath}
This has a natural interpretation as the ``worst-case'' RMS relative error in the dirty image due to gridding taken over all possible sets of visibilities on the given baselines.

The corresponding approximation to $\mathbf{A}$ is $\widetilde{\mathbf{A}},$ which is the adjoint of $\widetilde{\mathbf{A}}^\dagger$. This operates on $f(x)$ to give the degridded visibilities
\begin{linenomath}
\begin{equation}
\widetilde{V}_k = l_\mathrm{max}\sum_{r\in\mathbb{Z}} C(r - u'_k)\frac{1}{x_0}\int_{-x_0}^{x_0}
\exp(-i2\pi rx) h(x)f(x)\,\mathrm{d}x
\label{eq:adj_operator_1d}
\end{equation}
\end{linenomath}
because $h(x)$ is real and $C$ is even. For a particular $k$, there are only $W$ non-zero terms in the sum over $r$ because of the finite support of $C$. The integral over $x$ is evaluated only at integer values of $r$, and would in practice be done using an FFT. This process is summarized in Algorithm 2.

With this definition for degridding, we see that
\begin{linenomath}
\begin{align}
\left\lVert\mathbf{A}-\widetilde{\mathbf{A}}\right\rVert
&=\sup_{\mathcal{\mathbf{f}}\neq 0}
\frac
{\displaystyle\left(\frac{1}{M}\sum_{k=1}^M|{V}_k - \widetilde{V}_k|^2\right)^\frac{1}{2}}
{\displaystyle\left(\frac{1}{2x_0}\int_{-x_0}^{x_0} |f(x)|^2\,\mathrm{d}x \right)^\frac{1}{2}}
\label{eq:norm_diff_2}
\end{align}
\end{linenomath}
which has a natural interpretation as the ``worst-case'' RMS relative degridding error taken over all possible models of sky brightness on the region of interest $[-l_\mathrm{max},l_\mathrm{max}]$ which corresponds to normalized map coordinates $[-x_0, x_0]$.

%\color{green}
The operator norm provides a convenient way of placing bounds on the action of a linear operator, and is commonly used to quantify approximation errors. An application to the approximation of gridding kernels for $w$-projection is given by \citet{2019ApJ...874..174P}.
\color{black}

\subsection{Common error bounds for the approximation accuracy}

Equation (\ref{eq:equal_norms}) states that (\ref{eq:norm_diff_1}) and (\ref{eq:norm_diff_2}) are equal. We can use (\ref{eq:dirty_1d}) and (\ref{eq:dirty_1d_approx}) to derive an upper bound for both of these RMS relative errors. For a given set of visibilities ${V}_k$,
\begin{linenomath}
\begin{align}
|d(x)-\widetilde{d}(x)| &= \left|
\frac{1}{M}\sum_{k=1}^M {V}_k \left(e^{i2\pi u'_kx} - h(x)\sum_{r\in\mathbb{Z}} C(r-u'_k)e^{i2\pi rx}\right)
\right|
\label{eq:csum1}
\end{align}
\end{linenomath}
By using the Cauchy-Schwarz inequality, it follows that
\begin{linenomath}
\begin{equation}
|d(x)-\widetilde{d}(x)| \leq \left(\frac{1}{M}\sum_{k=1}^M|{V}_k|^2\right)^\frac{1}{2}
\left(\frac{1}{M}\sum_{k=1}^M|L(x,\nu_k)|^2\right)^\frac{1}{2}
\end{equation}
\end{linenomath}
where
\begin{linenomath}
\begin{equation}
L(x,\nu) = 1 - h(x)\sum_{s=-(W-1)/2}^{(W-1)/2}C(s-\nu)\exp{i2\pi(s-\nu)x}
\label{eq:csum2}
\end{equation}
\end{linenomath}
and, for each $k$, $\nu_k$ is defined such that the $W$ non-zero terms in the sum over $r$ in (\ref{eq:csum1}) coincide with those in the sum over $s$ in (\ref{eq:csum2}). Explicitly, $\nu_k=u'_k+(W-1)/2-\left\lfloor u'_k+W/2\right\rfloor$, and so $-\frac{1}{2}\leq\nu_k<-\frac{1}{2}$ for all $k$. Upon substituting these results into (\ref{eq:norm_diff_1}), we find that
\begin{linenomath}
\begin{equation}
\left\lVert\mathbf{A}-\widetilde{\mathbf{A}}\right\rVert =
\left\lVert\mathbf{A}^\dagger-\widetilde{\mathbf{A}}^\dagger\right\rVert \leq 2l_\mathrm{max}
\sqrt{\frac{1}{2x_0}\int_{-x_0}^{x_0}\frac{1}{M}\sum_{k=1}^{M}|L(x,\nu_k)|^2\,\mathrm{d}x}
\end{equation}
\end{linenomath}
For large $M$ and almost all realistic distribution of baselines, the values of $u'_k$ may be assumed to be uniformly distributed between the gridpoints of visibility space, so that $\nu_k$ are uniformly distributed within $\left[-\frac{1}{2},\frac{1}{2}\right]$. To good approximation, then, an upper bound for both norms is $2l_\mathrm{max}\mathcal{E}$, where
\begin{linenomath}
\begin{equation}
\mathcal{E}=\sqrt{\frac{1}{2x_0}\int_{-x_0}^{x_0}\int_{-\frac{1}{2}}^{\frac{1}{2}} |L(x,\nu)|^2\,\mathrm{d}\nu\mathrm{d}x}
\label{eq:norm_bound}
\end{equation}
\end{linenomath}
It is also useful to write
\begin{linenomath}
\begin{equation}
\mathcal{E}=\sqrt{\frac{1}{2x_0}\int_{-x_0}^{x_0}\ell(x)\,\mathrm{d}x},
\text{ where }
\ell(x) = \int_{-\frac{1}{2}}^{\frac{1}{2}}|L(x,\nu)|^2\,\mathrm{d}\nu
\end{equation}
\end{linenomath}
The map error function $\ell(x)$ was introduced by \citet{2020MNRAS.491.1146Y}, in which graphs were presented for a variety of gridding convolution functions in common use. A family of least-misfit gridding functions was also derived to minimize $\mathcal{E}$ for chosen values of $W$ and $x_0$. These may be used to calculate a common upper bound for the norms of the operator approximations, which give the RMS relative errors in degridding and gridding.

\vspace{1mm}
\begin{framed}
\noindent\textbf{Making a map with pixel size $\Delta l$ using an $N'$ point IFFT}
\begin{enumerate}[label={\arabic*}),leftmargin=1\parindent]
    \item Require visibility data on gridpoints $\Delta u=(N'\Delta l)^{-1}$ apart.
    \item Define $u'=u/(\Delta u)$, $x=l/(N'\Delta l)$. Then:
    \begin{enumerate}[label=(\alph*),leftmargin=2\parindent]
        \item $ul = u'x$, so $u'$ and $x$ are also conjugate
        \item $u'\in\mathbb{Z}$ at gridpoints of visibility space
        \item $-0.5\leq x < 0.5$ over the full extent of IFFT output
    \end{enumerate}
\end{enumerate}
\noindent For a given \emph{width} $W\in\mathbb{Z}^+$, and \emph{image cropping parameter}
$x_0\leq 0.5$, the \emph{least misfit gridding function} is an even function $C(u')$ which vanishes outside $[-W/2, W/2]$. It has an associated \emph{gridding correction function} $h(x)$ such that the following convolutional gridding procedure computes the dirty map on $N=2x_0N'$ pixels within $|x|\leq x_0$ with optimal accuracy:

\begin{enumerate}[label=({\arabic*}),leftmargin=1\parindent]
	\item Convolve each visibility $V_k$ at $u'_k$ with the least-misfit gridding function by adding the value $V_k C(r - u'_k)$ to the $W$ integer gridpoints $r\in\mathbb{Z}$ for which $C(r - u'_k)$ is non-zero.
	\item Apply the $N'$-point inverse fast Fourier transform (IFFT) to the gridded visibilities.
	\item Crop the IFFT image to retain the central $N = 2x_0 N'$ points for which $|x|\leq x_0$.
	\item Multiply the retained points by the gridding correction function $h(x)$ to give the dirty image.
\end{enumerate}
\end{framed}
\noindent \textbf{Algorithm 1.} Calculating an approximate naturally-weighted dirty map using `least-misfit gridding' in one dimension.
\vspace{1mm}

\vspace{1mm}
\begin{framed}
\noindent\textbf{Calculating model visibilities via FFT and degridding}\linebreak
The following procedure is the adjoint of Algorithm 1, provided that the same values of $x_0$ and $W$, the same functions $C(u')$ and $h(x)$ and the same number of points in the FFT are used.

\begin{enumerate}[label=({\arabic*}),leftmargin=1\parindent]
    \item Starting with a model image with $N$ points and pixel size $\Delta l$, zero pad the image to size $N'=N/(2x_0)$ pixels.
    \item Multiply the padded model image by the gridding correction function $h(x)$, where $x=l/(N'\Delta l).$
    \item Calculate the $N'$-point fast Fourier transform (FFT) to obtain values at integer $r\in\{-N'/2,...,N'/2-1\}.$
    \item To find the visibility at $u'_k$, multiply the result of the FFT at the gridpoint $r$ by $C(r-u'_k)$ and sum over the $W$ values of $r$ for which $C$ is non-zero.
\end{enumerate}
\end{framed}
\noindent \textbf{Algorithm 2.} Calculating visibilities from an image model using `least misfit degridding' in one dimension.
\vspace{1mm}
\color{black}

\section{Wide--field imaging methods}\label{s:imaging}

To deal with the complex $w$-term, the first method proposed was the three-dimensional Fourier transform method \citep{1989ASPC....6..259P}. This is mathematically straightforward and analytically precise, but it has been used only sparingly because of its high computational cost and the complex deconvolution process. In the same paper the polyhedron imaging method, or image-plane faceting, was introduced as an alternative with a much lower computational cost. The idea is to divide the entire wide FoV into several small FoVs, or {\it facets}, in each of which the two-dimensional Fourier relation is still applicable \citep{1989ASPC....6..259P}. The desired wide-field image can then be obtained by stitching the reconstructed images together. This method was implemented in the software \texttt{AIPS} \citep{1985daa..conf..195W} and has been used ever since. As an example, \citet{2017A&A...598A..78I} used the polyhedron method to make dirty images from the GMRT 150 MHz all-sky radio survey.

To take advantage of the two-dimensional Fourier transform without partitioning the FoV, algorithms such as the W-projection method \citep{Cornwell2003W, 2005ASPC..347...86C}, the W-snapshot method \citep{2012SPIE.8500E..0LC} and the W-stacking method \citep{humphreys2011analysis, 2014MNRAS.444..606O} have been proposed, as well as compressed sensing techniques \citep{2011MNRAS.413.1318M}.

The W-stacking method proposed in \citet{humphreys2011analysis} and \citet{2014MNRAS.444..606O} may be understood as calculating an approximation to the dirty image $I_D(l,m)$ associated with Equation \ref{eq:fourier_visibility}. For a set of $M$ visibilities $V(u_k, v_k, w_k)$, the dirty image satisfies
\begin{linenomath}
\begin{equation*}
	\frac{I_D(l,m)}{n} \propto \sum_k
	V(u_k,v_k,w_k)\exp[i2\pi(u_k l+v_k m+w_k(n-1))]
\end{equation*}
\end{linenomath}
where $n=\sqrt{1-l^2-m^2}$

In the W-stacking method, the visibilities are binned according to the value of $w_k$.
The interval $[w_\mathrm{min}, w_\mathrm{max}]$ within which all of the $w_k$ fall is divided into $N_w$ equal sub-intervals of width $\Delta w$, with midpoints $w[0]$ up to $w[N_w-1]$.
Let $S[t] = \{k: w[t]-\Delta w/2\leq w_k < w[t]+\Delta w/2\}$ denote those visibilities for which the nearest midpoint is $w[t]$. Then the approximation to the dirty map made is

\begin{linenomath}
\begin{multline}
	\frac{\widetilde{I_D}(l,m)}{n} \propto \sum_{t=0}^{N_w-1}
    \exp\{i2\pi w[t](n-1)\} \\
    \sum_{k\in S[t]}V(u_k,v_k,w_k)\exp\left[i2\pi(u_k l+v_k m)\right]
\label{eq: w_stacking}
\end{multline}
\end{linenomath}
\color{black}
The dirty map for each bin is computed using an FFT in conjunction with traditional two-dimensional convolutional gridding and gridding correction.

The resulting dirty image is then used to reconstruct the true sky brightness using a selected deconvolution algorithm.

Following \citet{2014MNRAS.444..606O}, the number of $w$-bins $N_w$ is determined by
\begin{linenomath}
\begin{align}\label{eq:N_w}
	N_w &\geq 2\pi\max_{l,m}(1-\sqrt{1-l^2-m^2})(w_{\rm max} - w_{\rm min})\nonumber \\
	&=2\pi(n_\mathrm{max}-n_\mathrm{min})(w_{\rm max} - w_{\rm min})
\end{align}
\end{linenomath}

A similar estimate for the number of $w$-bins, and an alternative
derivation for a given error tolerance, is also presented by \citet{2020PASA...37...41P}.

The W-stacking method can be summarized in the following steps:
\vspace{1mm}
\begin{framed}
\begin{enumerate}[label={\arabic*}]
	\item Determine the number of $w$-bins, $N_w$.
	\item Assign visibilities to these bins according to their $w$ values, using the nearest-neighbour rule.
	\item For each $w$-bin, the $w$ coordinate of the visibility is effectively replaced by $w[t]\equiv w_\mathrm{min} + (t+0.5)(w_\mathrm{max}-w_\mathrm{min})/N_w$,
	\begin{enumerate}
		\item Use a two-dimensional convolutional gridding method to reassign visibilities in this $w$-bin onto uniformly distributed grid points\label{step:gridding}
		\item apply a two-dimensional inverse FFT to the gridded visibilities,
		\item after cropping the undesired outer portion of the image, apply the phase shift for this bin by multiplication with $\exp\{i2\pi w[t](n-1)\}$.
		\label{step:oversampling}
	\end{enumerate}
	\item Sum the results for all $N_w$ $w$-layers
	\item Apply the gridding correcting function to the sum.
\end{enumerate}
\end{framed}
\noindent \textbf{Algorithm 3}. Original W-stacking method
\vspace{1mm}

The adjoint of the process of computing dirty maps may be used to calculate visibilities from a candidate model of the sky by means of the following steps. Because of the binning, a visibility with a baseline $(u_k,v_k,w_k)$ is computed as if it were measured at $(u_k,v_k,w[t])$, where $w[t]$ is the center of the nearest neighbouring bin to $w_k$.

\vspace{1mm}
\begin{framed}
\begin{enumerate}[label={\arabic*}]
	\item Determine the number of $w$ layers, $N_w$.
	\item For each $w$-bin with bin center $w[t]\equiv w_\mathrm{min} + (t-0.5)(w_\mathrm{max}-w_\mathrm{min})/N_w$,
	\begin{enumerate}
		\item Multiply the image model by the correcting function, zero-pad to the size of the FFT grid and apply the phase rotation,
		$\exp\{-i2\pi w[t](n-1)\}$
		\item apply the two-dimensional FFT to the corrected and phase-rotated image model,
		\item degrid by correlating with the chosen gridding function (this is equivalent to convolution, since the gridding function is even) in two dimensions so as to obtain the visibility at $(u_k, v_k, w[t])$.
	\end{enumerate}
\end{enumerate}
\end{framed}
\noindent \textbf{Algorithm 4.} Degridding procedure in the original W-stacking method
\vspace{1mm}

\citet{2014MNRAS.444..606O} pointed out that the value of $N_w$ has a great impact on the imaging quality, because errors have been introduced by assigning visibilities to the center of the nearest $w$-bin; these errors may be reduced by increasing $N_w$ at the cost of extra FFT computation. We shall assess the resulting inaccuracies after describing an improvement to W-stacking which involves gridding and degridding in all three dimensions.

\section{An Improved W-stacking Method}\label{sec:W-stacking}

\begin{figure}
    \centering
    \includegraphics[width=0.8\columnwidth]{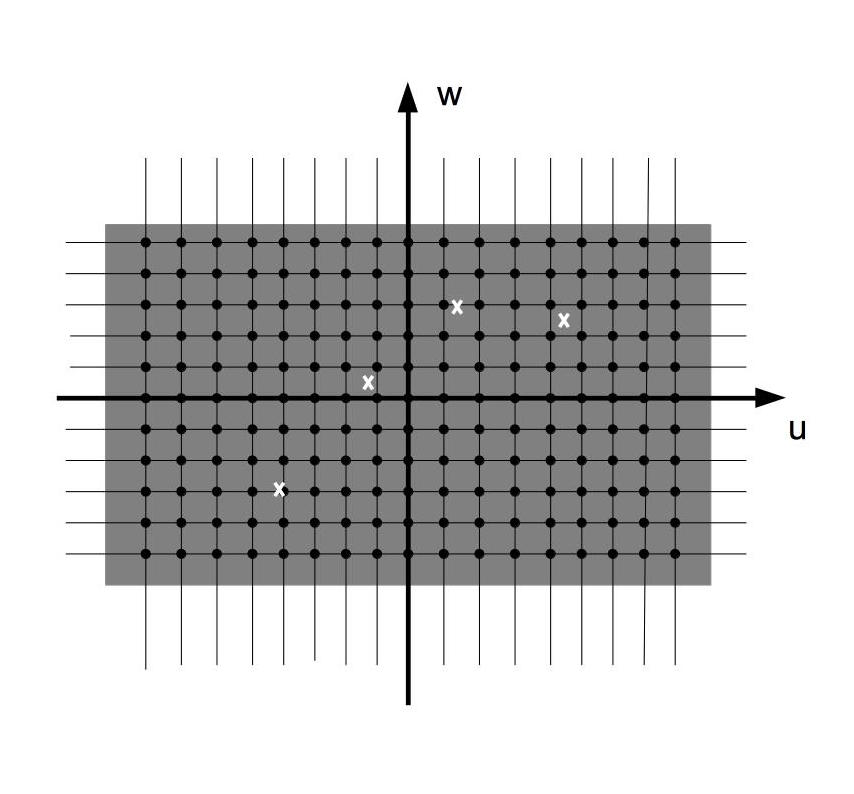}
    \caption{On the $w$-$u$ plane shown, the visibilities are to be gridded onto Cartesian grids represented by the black points. The $v$-axis is not shown, for simplicity. Hence, instead of a two-dimensional $uv$-plane gridding on each $w$-plane, the visibilities will be gridded onto three-dimensional grids.}
    \label{fig:w_stacking_3d}
\end{figure}

We now derive a significantly more accurate way of calculating the dirty image using a technique based on convolutional gridding for wide-field imaging. As explained in section \ref{sec:error_bounds}, its adjoint immediately provides an algorithm for calculating visibilities from an image model with essentially identical RMS normalized error bounds for both processes. This leads to a modification of the W-stacking algorithm which requires that visibilities be gridded in all three dimensions, rather than simply being moved to the nearest $w$-plane. As shown in Figure \ref{fig:w_stacking_3d} on the $w-u$ plane, the visibilities will be gridded onto rectangular grids represented by the black points. Hence, instead of a two-dimensional $uv$-plane gridding on each $w$-plane, the visibilities will be gridded onto genuinely three-dimensional grids. Gridding correcting functions must also be applied after the inverse Fourier transforms have been performed in order to compensate for the effects of this three-dimensional gridding operation.

\subsection{Normalized image coordinates and the celestial sphere}

Suppose that we are imaging a region of the celestial sphere lying in the range $l_{\rm{min}}\leq l \leq l_{\rm{max}}$ and $m_{\rm{min}}\leq m \leq m_{\rm{max}}$. For simplicity, take the centre of the map to be $l=m=0$, so that $l_{\rm{min}} = -l_{\rm{max}}$ and
$m_{\rm{min}} = -m_{\rm{max}}$. These limits correspond to the final size of the map after: carrying out convolutional gridding, taking the FFT, cropping the result and applying the gridding correction; in terms of the normalized map coordinates $(x,y)$, $x=\pm x_0$ corresponds to $l=\pm l_{\rm{max}}$ and $y=\pm y_0$ corresponds to $m=\pm m_{\rm{max}}$. We refer to $x_0$ and $y_0$ as the ``image cropping parameters'', because they define the fraction of the FFT output retained in each direction. The extent of the FFT grid expressed in normalized coordinates is $[-0.5,0.5]\times [-0.5,0.5]$, so for example the choice $x_0=y_0=0.25$ corresponds to cropping the result of the FFT to half of its size in both directions.

\begin{figure}
    \centering
    \includegraphics[width=\columnwidth]{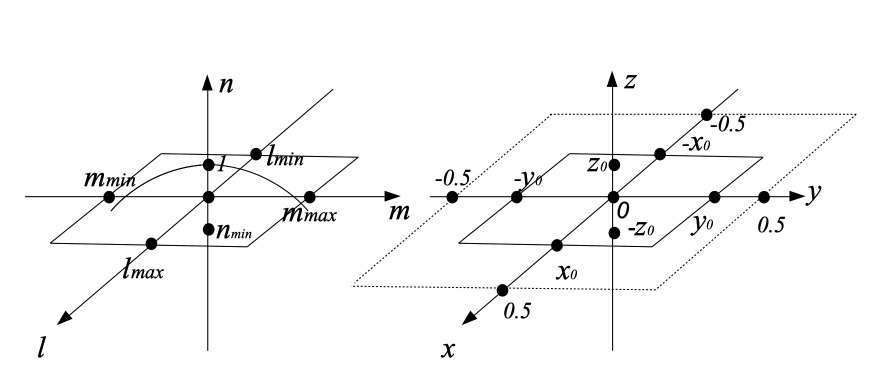}
    \caption{The three-dimensional coordinate system formed by the direction cosines $(l,m,n)$ (left) and the normalised coordinate system $(x,y,z)$ (right).}\label{fig:x0_coordinate_3d}
\end{figure}

As shown in Figure \ref{fig:x0_coordinate_3d}, the normalized image coordinates are related to $(l,m)$ via the relations $x=x_0 l/l_{\rm{max}}$ and $y=y_0 m/m_{\rm{max}}$. In the $z$-direction it is convenient to write $z=(n-n_0)/n_{\rm{scale}}$, where $n_0$ is a quantity which offsets the plane $z=0$ from the tangent plane to the celestial sphere. An appropriate choice of $n_0$ and the scaling factor $n_{\rm{scale}}$ will be stated below.

The minimum and maximum values of $n$ over the region of interest of the map (from $n_\mathrm{min}=\sqrt{1-l^2_\mathrm{max}-m^2_\mathrm{max}}$ to $n_\mathrm{max}=1$ in the figure) will be mapped to $z=-z_0$ and to $z=z_0$ respectively, where $z_0$ plays an analogous role to $x_0$ when we consider gridding in that direction.

\subsection{High-accuracy approximation of the operators for modelling of visibilities and mapping in wide-field imaging}

In terms of the normalized coordinates, equation (\ref{eq:fourier_visibility}) may be written as
\begin{linenomath}
\begin{multline}
{V}_k = \frac{l_{\rm{max}}m_{\rm{max}}}{x_0 y_0}\exp[-i2\pi w_k(n_0-1)]\times\\
\int_{-x_0}^{x_0}\int_{-y_0}^{y_0}
f(x,y)\exp[-i2\pi (u'_k x + v'_k y + w'_k z)]\,\mathrm{d}x\mathrm{d}y
\end{multline}
\end{linenomath}
where $f(x,y)=I(x l_\mathrm{max}/x_0,y m_\mathrm{max}/y_0)/\sqrt{1-l^2-m^2}$ and
the normalized visibility coordinates conjugate to $x$, $y$ and $z$ are $u'=l_\mathrm{max}u/x_0,$ $v'=m_\mathrm{max}v/y_0,$ and $w'=n_\mathrm{scale}w$.
We have also used the fact that $w_k (n-1) = w_k n_\mathrm{scale} z + w_k(n_0-1)$. This relation is the analogue of $\mathbf{V}=\mathbf{A}\mathbf{f}$ in (\ref{eq:fwd_1d_op}).

The naturally-weighted dirty image for a collection of visibilities ${V}_k\equiv{V}(u'_k,v'_k,w'_k)$ is associated with the adjoint operator $\mathbf{A}^\dagger$ taken with respect to inner products defined analogously to (\ref{eq:image1d_ip}) and (\ref{eq:natural_vis_ip}). We find that $\mathbf{A}^\dagger\mathbf{V}(x)=4l_\mathrm{max}m_\mathrm{max} d(x,y)$, where
\begin{linenomath}
\begin{equation}
d(x,y) = \frac{1}{M}\sum_{k=1}^M \left[{V}_k \mathrm{e}^{i2\pi w_k (n_0-1)}\right]\mathrm{e}^{i2\pi w'_k z} \mathrm{e}^{i2\pi v'_k y} \mathrm{e}^{i2\pi u'_k x}
\end{equation}
\end{linenomath}
Despite the formal similarity to a three-dimensional discrete inverse Fourier transform, this is evaluated on a two-dimensional surface, because $z$ is determined by $x$ and $y$, via
\begin{linenomath}
\begin{equation}\label{eq:z-on-sphere}
z=\frac{\sqrt{1 - l^2 - m^2} - n_0}{n_{\rm{scale}}},\;\text{where}\; l=\frac{l_{\rm{max}}x}{x_0}\;\text{and}\;m=\frac{m_{\rm{max}}y}{y_0}.
\end{equation}
\end{linenomath}

We turn now to the approximation of the operator $\mathbf{A}^\dagger$ by $\widetilde{\mathbf{A}}^\dagger$ which arises through the use of convolutional gridding.
As discussed above, in each dimension this allows a discrete (inverse) Fourier transform of data $F_k$ located at non-uniformly spaced points $u_k$ defined by
\begin{linenomath}
\begin{equation}
f(x) = \sum_k F_k \exp(i2\pi u'_k x)
\label{eqn:gridding1}
\end{equation}
\end{linenomath}
to be calculated accurately within an interval $|x|\leq x_0$ using
\begin{linenomath}
\begin{equation}\label{eqn:gridding2}
f(x)\approx h(x)\sum_{r\in\mathbb{Z}}\left[ \sum_k F_k C(r-u'_k) \right]\exp(i2\pi rx)
\end{equation}
\end{linenomath}
Although the sum over $r$ runs formally over all the integers, there are only a finite number of non-zero terms because $C$ has finite support and the values $u_k$ lie within a bounded range. If the FFT is to be used to evaluate the sum, $N$ must be chosen to include all non-zero terms.

By making multiple use of the approximate equality between the right-hand sides of (\ref{eqn:gridding1}) and (\ref{eqn:gridding2}) --- once for each of the three dimensions --- and allowing for the possibility of using different gridding functions in each, the gridded dirty map is
\begin{linenomath}
\begin{equation}
\label{eq:gridded-dirty-map}
\widetilde{d}(x,y)=
h_x(x)h_y(y)h_z(z)\sum_{t\in\mathbb{Z}} \left[\sum_{r\in\mathbb{Z}}\sum_{s\in\mathbb{Z}}
G_{rst}\mathrm{e}^{i2\pi r x}
\mathrm{e}^{i2\pi s y}
\right]\mathrm{e}^{i2\pi t z}
\end{equation}
\end{linenomath}
where
\begin{linenomath}
\begin{equation}
\label{eq:gridded_vis}
G_{rst}=
\frac{1}{M}\sum_k \left[{V}_k \mathrm{e}^{i2\pi w_k (n_0-1)} \right] C_x(r-u'_k)  C_y(s-v'_k) C_z(t-w'_k).
\end{equation}
\end{linenomath}
This result indicates that the weighted visibilities should be multiplied by the phase factor $\exp[i2\pi w_k(n_0 -1)]$ and convolved in all three directions using the separable product $C_x(u')C_y(v')C_z(w')$ of gridding functions in each direction. The sums over $r$ and $s$ can be performed using the complex two-dimensional inverse FFT algorithm, since the values of $x$ and $y$ are required on uniformly spaced grids. As in conventional W-stacking, we have a collection of such 2D IFFTs, one for each value of $t$.

The numbers of points $N'_x\times N'_y$ for the two-dimensional FFTs are chosen so that they cover the non-zero terms in the sums over $r$ and $s$ in (\ref{eq:gridded-dirty-map}). This leads to the lower bounds $N'_x> (u_{\rm{max}} - u_{\rm{min}})l_{\rm{max}}/x_0+W_x$ and $N'_y> (v_{\rm{max}} - v_{\rm{min}})m_{\rm{max}}/y_0+W_y$, where $u$ and $v$ are measured in wavelengths and $W_x$ and $W_y$ define the widths of the gridding functions.

\subsection{Combining images from the $w$ layers and determining the number of layers required}

Since the values of $z$ associated with points $(x,y)$ on the image do not fall on a regular grid, the sum over $t$ cannot be evaluated using an inverse FFT algorithm. Instead, (\ref{eq:z-on-sphere}) is used to find the phase factors $\exp(i2\pi t z(x,y))$ which are multiplied by the results of the two-dimensional inverse FFTs before they are summed up.

An important feature of the calculation is that gridding correction involves the product of three terms $h_x(x)h_y(y)h_z(z(x,y))$, and that $z(x,y)$ is given by equation \ref{eq:z-on-sphere}. Inclusion of the factor in the $z$ direction is critical to the accuracy of the method. This product is computed once for all points $(x,y)$ which comprise the final image, and is applied to the sum of the stack.
Accurate calculation of $h(z)$ may be performed using polynomial interpolation of high degree within a table of $h$ specified initially on a uniform grid.

It remains finally to determine the values of $n_0$ and $n_{\rm{scale}}$, which will also determine the number of $w$ layers in the $w$-direction (i.e., the number of non-zero terms in the sum over $t$). If we use optimal gridding convolution functions in all three directions, the results will be accurate within $|x|\leq x_0$, $|y|\leq y_0$ and $|z|\leq z_0$. From (\ref{eq:z-on-sphere}), we see that $z$ is a maximum at the map center where $x=y=0$, and is a minimum at the corners of the map where $x=x_0, y=y_0.$ In particular,
\begin{linenomath}
$$z_{\rm{max}} = \frac{1-n_0}{n_{\rm{scale}}}\quad\text{and}\quad
z_{\rm{min}} =
\frac{\sqrt{1 - l_{\rm{max}}^2 - m_{\rm{max}}^2} - n_0}{n_{\rm{scale}}}$$
\end{linenomath}
To use the region of accuracy of the gridding convolution function in the most sensible way, we choose $z_{\rm{min}}=-z_0$ and $z_{\rm{max}}=+z_0$. This leads to
\begin{linenomath}
\begin{equation}\label{eq:n_params}
n_{\rm{scale}} = \frac{1-\sqrt{1 - l_{\rm{max}}^2 - m_{\rm{max}}^2}}{2z_0}\;\text{and}\; n_0=1- z_0{n_{\rm{scale}}}
\end{equation}
\end{linenomath}
which are the promised expressions for the quantities introduced above.

Corresponding to the normalized coordinates $z_{\rm{min}}$ and $z_{\rm{max}}$, the map extends from $n_{\rm{min}} = \sqrt{1 - l_{\rm{max}}^2 - m_{\rm{max
}}^2}$ to $n_{\rm{max}} = 1$. Written in terms of these quantities, (\ref{eq:n_params}) becomes
\begin{linenomath}
\begin{equation}\label{eq:n_params_alt}
n_{\rm{scale}} = \frac{n_{\rm{max}}-n_{\rm{min}}}{2z_0}\quad\text{and}\quad n_0=\frac{n_{\rm{max}}+n_{\rm{min}}}{2}
\end{equation}
\end{linenomath}

The separation between two adjacent $w$ layers in the $w$-direction is $1/n_{\rm{scale}}$ wavelengths, and so the number of $w$ layers required must satisfy
\begin{equation}\label{eq:N_{w'}+W}
N_{w'} > {\frac{(n_\mathrm{max}-n_\mathrm{min})(w_\mathrm{max}-w_\mathrm{min})}{2z_0}} + W_z
\end{equation}
where the further $w$ layers are necessary to allow for the support of the gridding function.%, \textcolor{orange}{especially for visibilities which are located close to the top or bottom $w$-plane in the $w$-direction}

The above choice of $z_\mathrm{min}$ and $z_\mathrm{max}$ minimizes the number of $w$-layers required, by utilizing the full range $[-z_0, z_0]$ over which the convolutional gridding method has been optimized. By setting $z_{\rm{max}}=+z_0$, however, the centre of the map $x=y=0$ is mapped to a point at which $h_z(z),$ the gridding function in the $z$ direction, is a maximum. Since the map error associated with the use of the least-misfit optimal gridding functions \citep{2020MNRAS.491.1146Y} is very small and is fairly uniform over the entire interval $[-z_0,z_0]$, this procedure is usually acceptable, but the discussion around Figure~\ref{fig:Wstacking_correction} below should be kept in mind.

If it is desired to place the minimum of the gridding correction at the map center, an alternative choice is to take $z_\mathrm{max} = 0$ and $z_\mathrm{min} = -z_0$, leading to
\begin{linenomath}
\begin{equation}\label{eq:n_params_no_offset}
n_{\rm{scale}} = \frac{1-\sqrt{1 - l_{\rm{max}}^2 - m_{\rm{max}}^2}}{z_0}\;\text{and}\; n_0=1
\end{equation}
\end{linenomath}
in place of (\ref{eq:n_params}). We refer to this variant of the algorithm as being without $z$ offset because, with this choice in Figure \ref{fig:x0_coordinate_3d}, the plane $z=0$ is tangential to the celestial sphere at the phase centre (where $n=1$), rather than being offset to $(n_\mathrm{max}+n_\mathrm{min})/2$. The number of $w$-layers needed for this variant is such that
\begin{equation}
N_{w''} > {\frac{(n_\mathrm{max}-n_\mathrm{min})(w_\mathrm{max}-w_\mathrm{min})}{z_0}} + W_z
\end{equation}

It is of interest to compare the number of layers required with the number recommended for the original W-stacking algorithm (\ref{eq:N_w}).
If we neglect the additional $W_z$ term required for the full gridding and use a typical value of $z_0=0.25$, the new algorithm requires about a third
and the variant without $z$ offset requires about two-thirds the number of layers.

This reduces the cost of computing the FFTs, although the gain should be offset against the cost of the three-dimensional gridding. In section \ref{sec:performance} we shall see that, to achieve comparable accuracy for wide-field imaging, the number of layers required by the original W-stacking algorithm is often much greater than that recommended by (\ref{eq:N_w}), so that the cost of the FFT computations is more significant than that of 3-d gridding.
\color{black}

\subsection{Details for efficient implementation}

A special case occurs if we wish to map the entire hemisphere of the sky, for which we set $n_{\rm{scale}}=1/(2z_0)$ and $n_0=\frac{1}{2}$, since in this case $n_{\rm{min}}=0$ and $n_{\rm{max}}=1$.

In practice, since the sky brightness is real, a visbility $V_k$ at the point $(u_k,v_k,w_k)$ with $w_k<0$ may be replaced by its complex conjugate $V^*_k$ at the point $(-u_k,-v_k,-w_k)$, so that $w_{\rm{min}}$ can be taken as zero.

It is convenient to perform the convolution in the $w$-direction first, spreading each weighted visibility multiplied by the phase factor $\exp[i2\pi w_k(n_0-1)]$ over the $W_z$ $w$ layers as determined by the gridding function. The values assigned to each of the $N_{w'}$ $w$ layers are then gridded onto the $uv$ grids. An inverse 2-dimensional FFT is applied to each $w$-plane, followed by image cropping and application of the phase factor $\exp(i2\pi t z(x,y))$. Instead of saving the $N_{w'}$ images, we accumulate them and save their sum. This accumulation can be performed on the grid with the final (cropped) map size $N_x\times N_y$ so as to save memory. In regard to the size of the two-dimensional FFT, $N_x=2x_0N_x'$ and $N_y=2x_0N_y'$.

It is often desirable to use values of $N'_x$ and $N'_y$ which are greater than the lower bounds given in the above analysis by a factor $\gtrsim 5$, so that the Fourier transform performs a band-limited interpolation and gives a smooth map. Since the accuracy of mapping and the calculation of model visibilities depends primarily on the parameters of the gridding functions, rather than on the size of the FFTs, the accuracy is unaffected, although an image model with more pixels can be helpful for image reconstruction algorithms which necessitate optimization over other criteria (such as entropy) in addition to data misfit. Since image models are two-dimensional, there is no need to use more than the minimum number of $w$-layers in the $z$ direction given by the inequality (\ref{eq:N_{w'}+W}).

The steps required to calculate dirty images from measured visibilities via gridding, and to calculate visibilities from a model image via degridding using the improved W-stacking method, may be summarized as follows:

\vspace{1mm}
\begin{framed}
\begin{enumerate}[label={\arabic*}]
	\item Determine the number of $w$-layers, $N_{w'}$, and calculate $n_0$, $n_\mathrm{scale}$ using (\ref{eq:n_params}) and (\ref{eq:N_{w'}+W}).
	The $w$-layers are indexed by the integer $t$, where
	$w[t]=t/n_\mathrm{scale}$.
	
	\item Starting with $M$ measured visibilities, we grid in the $w$ direction by distributing $M W_z$ values among the $N_{w'}$ $w$-layers, so that
	conventional two-dimensional mapping can be performed for each layer.
	
	Each visibility $V_k$ on baseline $(u_k, v_k, w_k)$ is multiplied by $\exp[i2\pi w_k(n_0-1)]$. It is
	assigned to the $W_z$ layers with indices $t$ which satisfy $|w[t]-w_k|\leq W_z/(2n_\mathrm{scale})$
	after being multiplied by the gridding convolution function $C_z(n_\mathrm{scale}(w[t]-w_k))\equiv C_z(t-w'_k).$

    \item For each $w$-layer with index $t$,
	\begin{enumerate}
		\item Use two-dimensional convolutional gridding with functions $C_x$ and $C_y$
		to place the assigned visibilities onto uniformly distributed gridpoints at integer values of $r$ and $s$ in equation (\ref{eq:gridded_vis}).
		\item Compute the two-dimensional inverse FFT.
		\item Crop the outer portion of the image, retaining only the portion $|l|\leq l_\mathrm{max}$ and $|m|\leq m_\mathrm{max}$, or equivalently
		$|x|\leq x_0$ and $|y|\le y_0$.
		\item Apply the phase factor $\exp\{i2\pi w[t](n-n_0)\}$ or equivalently $\exp\{i2\pi t z(x,y)\}$, where $z(x,y)$ is given by (\ref{eq:z-on-sphere}).
	\end{enumerate}

    \item Accumulate the results of the previous step for all of the layers into an initially empty array.

    \item Multiply by the 3-dimensional gridding correction $h_x(x_0 l/l_\mathrm{max})h_y(y_0 m/m_\mathrm{max})h_z((n-n_0)/n_\mathrm{scale})$, or equivalently $h_x(x) h_y(y) h_z(z(x,y)).$

\end{enumerate}
\end{framed}
\noindent \textbf{Algorithm 5}. Generating dirty images via improved W-stacking method and 3-dimensional gridding
\vspace{1mm}

\vspace{1mm}
\begin{framed}
\begin{enumerate}[label={\arabic*}]
	\item Determine the number of $w$ layers, $N_{w'}$, and the value of $n_0$.
	\item Multiply the two-dimensional image model by the 3-dimensional correction function $h_x(x) h_y(y) h_z(z(x,y))$, where $z(x,y)$ is given by (\ref{eq:z-on-sphere}).
	\item For each $w$-plane with index $t$,
	\begin{enumerate}
		\item Multiply the corrected image model by the phase factor
		 $\exp[-i2\pi t z(x,y)].$
		\item Pad zeros around the phase-shifted image model, extending the ranges $|x|\leq x_0$ and $|y|\leq y_0$ to $|x|\leq 0.5$ and $|y|\leq 0.5$.
        \item Compute the two-dimensional FFT.
		\end{enumerate}
	\item Having obtained $N_{w'}$ sets of two-dimensional FFT results, perform a three-dimensional correlation in all three directions using the least-misfit gridding functions $C_x$, $C_y$ and $C_z$. This is equivalent to a convolution because the functions are even.
	\item For a visibility point at $w_k$, multiply the result by the phase factor $\exp[-i2\pi w_k(n_0-1)]$.
\end{enumerate}
\end{framed}
\noindent \textbf{Algorithm 6}. Calculation of model visibilities by the improved W-stacking method and 3-dimensional degridding
\vspace{1mm}

\section{Performance of the improved W-stacking method}\label{sec:performance}

\subsection{Quantifying computational cost for a specified accuracy}

We now compare the computational cost of the original and improved W-stacking methods for the problem of generating a dirty image of size $N_x\times N_y$ pixels extending over some portion of the celestial sphere by means of convolutional gridding of $N_v$ measured visibilities. The cost of calculating visibilities from a model image by degridding scales in the same way, although a straightforward implementation of the adjoint operator may require complex arithmetic where real arithmetic sufficed for the original operator, or vice versa.

Table \ref{tab:W-stacking_cost_comparison} summarizes the operation counts of the dominant terms in the computation. In the improved method, gridding is performed in three dimensions using convolutional functions with widths $W_x$, $W_y$ and $W_z$ optimized over the intervals $[-x_0,x_0]$, $[-y_0,y_0]$ and $[-z_0,z_0]$. For the original method, gridding is performed only in two dimensions, so that we may take $W_z=1$. The number of $w$ layers in the stack is denoted by $N_w$ for the original method, and $N_{w'}$ for the improved method. The two terms are respectively the cost of gridding and of computing the Fast Fourier transforms.

\begin{table}
\centering
 \caption{Overall computational cost of the original and improved W-stacking methods}
 \label{tab:W-stacking_cost_comparison}
 \begin{tabular}{lcc}
  \hline
  Method & Computational cost \\
  \hline
  Original & $\mathcal{O}(N_v W_x W_y) + \mathcal{O}\bigg(\frac{N_{w}N_x N_y}{4x_0 y_0}\log\big(\frac{N_x N_y}{4x_0 y_0}\big)\bigg)$\\
  Improved & $\mathcal{O}(N_v W_x W_y W_z) + \mathcal{O}\bigg(\frac{N_{w'}N_x N_y}{4x_0 y_0}\log\big(\frac{N_x N_y}{4x_0 y_0}\big)\bigg)$\\
  \hline
 \end{tabular}
\end{table}

The additional factor of $W_z$ in the cost of three-dimensional gridding is balanced against the number of $w$ layers required for calculation of the Fourier transforms. The dominant term in the number of layers required (see equation \ref{eq:N_{w'}+W}) is inversely proportional to $z_0$ and is directly proportional to the non-planarity of the sky ($n_\mathrm{max}-n_\mathrm{min}$) and of the measurement array along the line of sight ($w_\mathrm{max}-w_\mathrm{min}$). By calculating how the approximation error in the $z$ direction varies for various choices of $W_z$ and $z_0$, we are able to distribute the computation cost between the two terms.

%\subsection{Accuracy of the W-stacking methods}{\label{sec:W-stacking_accuracy}}

The original W-stacking method may be considered as a special case of the improved method when a top-hat gridding convolutional function is used with $W_z=1$. We therefore begin with an analysis of the improved method.

Parallel to the analysis in section \ref{sec:error_bounds}, the RMS relative errors associated with gridding and degridding are related and are given by
\begin{linenomath}
\begin{multline}
\left\lVert\mathbf{A}-\widetilde{\mathbf{A}}\right\rVert =
\left\lVert\mathbf{A}^\dagger-\widetilde{\mathbf{A}}^\dagger\right\rVert =
4l_\mathrm{max}m_\mathrm{max}\times\\
\sup_{\mathbf{V}\neq 0}
\frac{\displaystyle\left(\frac{1}{4x_0y_0}\int_{-x_0}^{x_0}\int_{-y_0}^{y_0} |d(x,y)-\widetilde{d}(x,y)|^2\,\mathrm{d}x\mathrm{d}y \right)^\frac{1}{2}}
{\displaystyle\left(\frac{1}{M}\sum_{k=1}^M|{V}_k|^2\right)^\frac{1}{2}}
\label{eq:norm_bound2}
    \end{multline}
\end{linenomath}
Calculation of $\widetilde{d}(x,y)$ involves a three-dimensional gridding process. An argument analogous to that leading to (\ref{eq:norm_bound}) shows that an upper bound on (\ref{eq:norm_bound2}) can be found, and that it can be minimized over the space of three-dimensional gridding convolution and correction functions. It may further be shown that the multi-dimensional optimal solutions separate into the product of optimal functions in each dimension, and that the overall RMS relative gridding or degridding error is approximately the quadrature sum of the errors in each dimension (in the small error limit).

In the $x$ and $y$ directions, the RMS gridding and degridding errors can be calculated from graphs of the map error function $\ell(x)$ given in \citet{2020MNRAS.491.1146Y}. For example, the least-misfit gridding function optimized for $x_0=0.25$ and $W=7$ achieves $\mathcal{E}=1.3\times 10^{-7}$, which is appropriate for single-precision processing. Unless  the error in the $z$ direction is chosen to be comparable to these values, it will be the dominant source of error in the algorithm.

In the $z$ direction, we recall that $\mathcal{E}$ depends on both $W_z$ and $z_0$, so that if we wish to retain the same accuracy while reducing the value of $W_z$, $z_0$ must also be reduced. From (\ref{eq:N_{w'}+W}), reducing the value of $z_0$ means that the number of layers $N_{w'}$ in the W-stack must be increased correspondingly, increasing the computational cost of performing the FFT.

\begin{figure}
    \centering
    \includegraphics[width=\columnwidth]{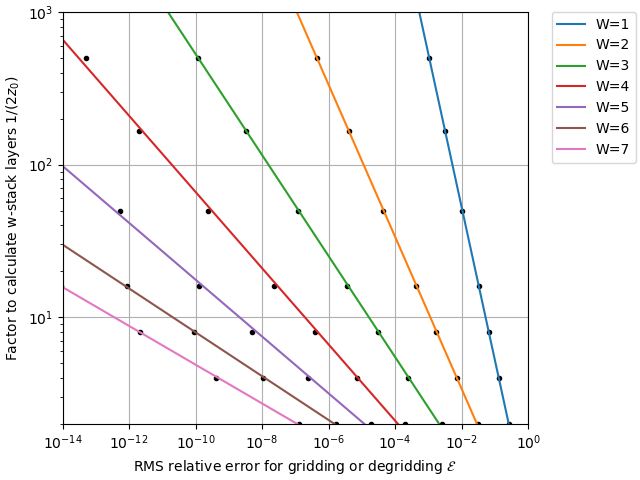}
    \caption{Factor for calculating the number of layers required in W-stack, using least-misfit gridding
    convolution functions of various widths $W$ in the $z$ direction to achieve a desired accuracy for making dirty images via gridding, or calculating model visibilities via degridding.}\label{fig:improved_wstack_accuracy}
\end{figure}

Figure \ref{fig:improved_wstack_accuracy} shows the trade-off between accuracy $\mathcal{E}$ and $1/(2z_0)$ for the family of least-misfit gridding functions having $W_z=1$ through $W_z=7$. From the value of $1/(2z_0)$, the number of layers in the W-stack $N_{w'}$ can be computed using $(\ref{eq:N_{w'}+W})$. Other choices of gridding functions will perform worse.

As an example, suppose we use $W_x=W_y=7$ and $x_0=y_0=0.25$ so that the RMS relative error is $\approx 10^{-7}$. Compared to using $W_z=7$, we see from a vertical line through $\mathcal{E}=10^{-7}$ that about $6$ times the number of layers if we use $W_z=4$, $27$ times the number of layers are needed if we use $W_z=3$, and about $500$ times the number of layers if we use $W_z=2$. When high accuracy is required, the cost of gridding is rapidly dominated by the FFTs needed for the large number of $w$-layers. If lower accuracy is acceptable, the cost of three-dimensional gridding can be reduced by using smaller values of $W_x$, $W_y$ and $W_z$ but the choice $W_z=1$ is almost never optimal unless the non-planar terms are negligible.

\color{black}
\subsection{Accuracy of the original W-stacking algorithm with $W_z=1$}

The original W-stacking method may be analyzed as the special case in which $W_z=1$ and the gridding convolution function is a top hat, i.e., $C_z(w')=1$ for $|w'| < 1/2$ and zero otherwise. The $W$-bins of the method then coincide with the $w$-layers of the general theory.
With a correction function of $h(z)=1/\sinc z$, (\ref{eq:csum2}) is evaluated as
\begin{linenomath}
\begin{equation}
L_z(z,\nu) = 1 - \frac{\exp(-i2\pi\nu z)}{\sinc z}
\end{equation}
\end{linenomath}
From this result it follows that $\mathcal{E}\approx \sqrt{\pi/3}\, z_0$ for small $z_0$. On Figure \ref{fig:improved_wstack_accuracy}, this yields a line that is visually indistinguishable from the result for $W=1$. This makes it clear that the original W-stacking method is capable of only modest accuracy, which improves slowly with the number of layers in the stack (since $\mathcal{E}\propto z_0\propto N_w^{-1}$). The error introduced by the effective nearest-neighbour gridding in the $z$ direction dominates that in the $x$ and $y$ directions for most choices of parameters in which the non-planar terms are important. To attain levels of accuracy comparable to those in the $x$ and $y$ directions, the number of $w$-layers required is typically several orders of magnitude larger than that recommended by (\ref{eq:N_w}).

\color{black}
With the improved W-stacking algorithm, the ability to use a larger value of $W_z$ allows the accuracy to improve more rapidly with the number of layers. It becomes feasible to divide the error budget more evenly between the gridding processes in all three dimensions even when mapping large areas of the sky, and to balance the cost of gridding and computing FFTs as necessary. For simplicity, we recommend using identical gridding convolution and correction functions in all three directions, i.e., $W_x=W_y=W_z$ and $x_0=y_0=z_0$, unless there are compelling reasons to do otherwise.

\subsection{Comparing Accuracy using Numerical Simulations}

We now present various illustrative examples, in which the difference between the exact dirty map calculated via a discrete Fourier transform and the approximate dirty map calculated via the original and improved W-stacking methods are compared. These comparisons also give an indication of the accuracy to which visibilities can be computed for a model sky using degridding, which is a key step of image reconstruction.

We synthesize data for a field consisting of 34 point sources, whose fluxes and positions are specified in the Appendix. The positions are given in terms of pixel numbers on a map where the inter-pixel separation in each direction is $24$ arcseconds. The final map is of size $900\times 900$ pixels, extending 6 degrees in each direction. The visibility data are calculated so as to simulate an observation on the baselines of the VLA B-array at $74$ MHz. The hour angle changes from $-5$h to $+5$h during the observation, and $2$ minutes of data are collected at intervals of $1$h, resulting in $22$ minutes of data. Since the plane of the array varies during the observation, the range of $w$-values is significant in these data.

An exact dirty image is calculated from the synthetic visibiity data using the slow Discrete Fourier Transform (DFT) for reference. Approximate dirty images are also calculated using the original and improved W-stacking methods for a variety of parameter values. The dirty image approximation error is the difference between the dirty image computed using each approximate method and the reference (exact) dirty image.

\begin{figure}
    a)\centering\includegraphics[width=\columnwidth]{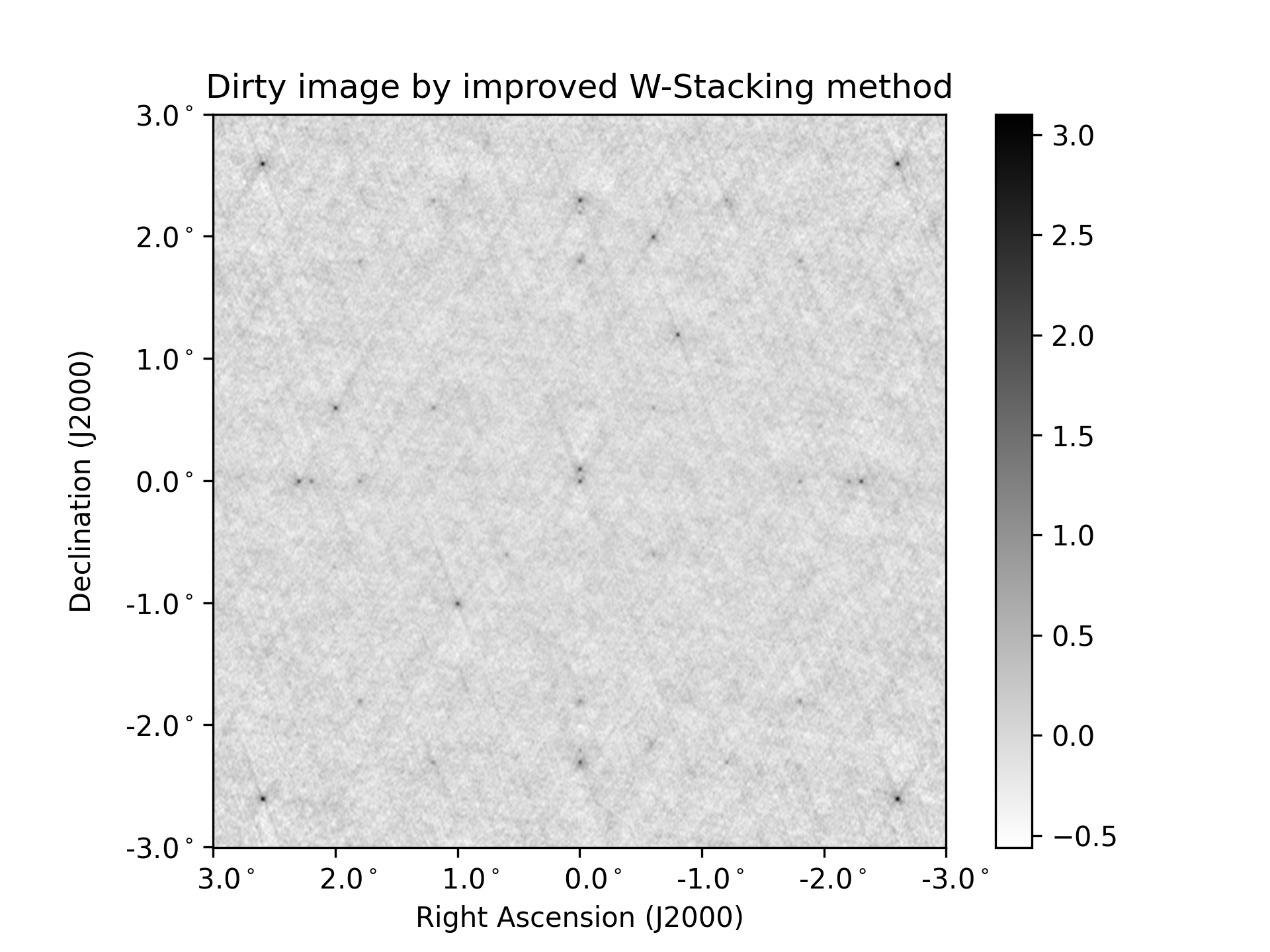}
    \hspace{1cm}
    b)\centering\includegraphics[width=\columnwidth]{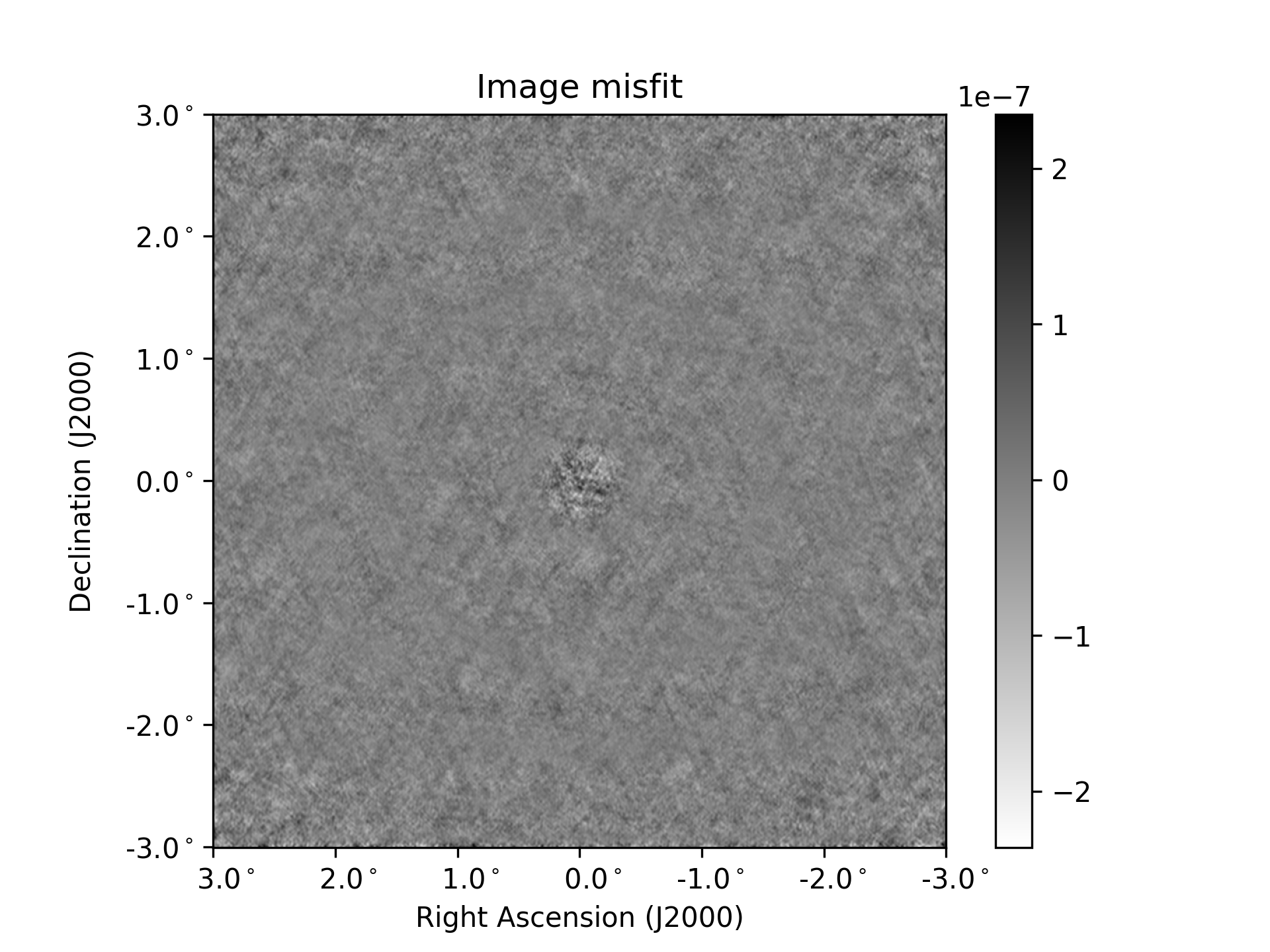}
    \caption{a) Dirty image constructed using the improved W-stacking method; b) its difference from the DFT dirty image. The least-misfit gridding function is used with $W_x=W_y=W_z=7$ and $x_0=y_0=z_0=0.25$}
    \label{fig:Wstacking_FFT}
\end{figure}

Figure~\ref{fig:Wstacking_FFT} shows the result for our improved W-stacking method with $W_x=W_y=W_z=7$ and $x_0=y_0=z_0=0.25$. The number of $w$-layers used is 22, in accordance with equation (\ref{eq:N_{w'}+W}). From Figure \ref{fig:Wstacking_FFT}b we see that, for this method, the dirty image approximation error is fairly uniform over the image and is at a low level, with an RMS value of $1.8 \times10^{-8}$.
%it is under W-stacking-improved Optimised with decreased w-planes (images produced for paper).ipynb

To compare the various methods, we find the RMS value of the dirty image approximation error. As the approximation error may vary substantially over the dirty image, we also compute the RMS error over concentric square regions which extend from $-R\leq x \leq R$ and $-R\leq y \leq R$, so that small values of $R$ correspond to considering only the center of the image, while $R=x_0=y_0$ corresponds to calculating the RMS error over the whole image.

\begin{figure}
\centering\includegraphics[width=\columnwidth]{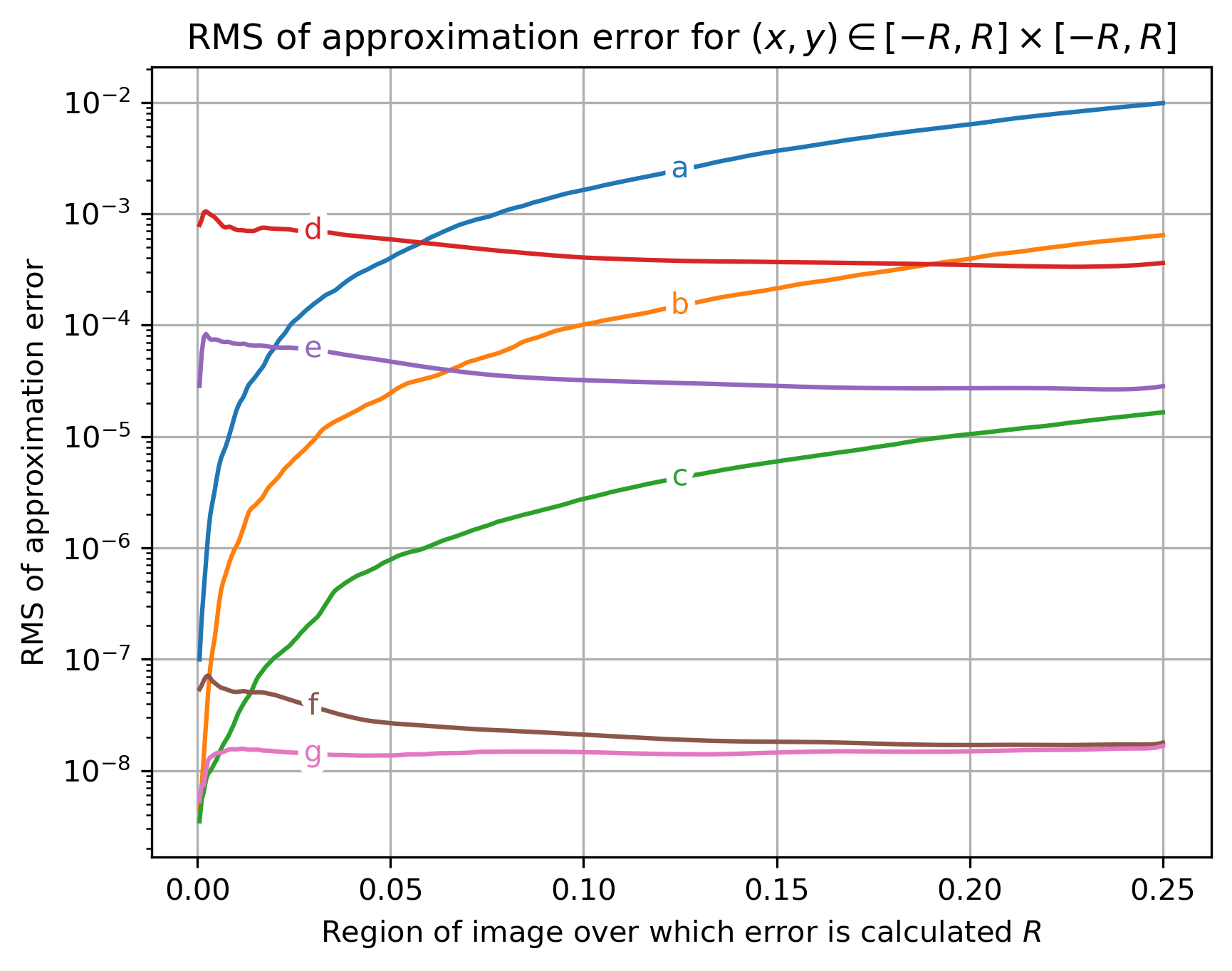}
\caption{The RMS of the dirty image approximation error calculated over the central region of the image $[-R,R]\times[-R,R]$ for various W-stacking methods. The lines are labelled as in Table \ref{tab:rms-approx-error}. Since $x_0=y_0=0.25$ in the cases shown in the figure, $R=0.25$ corresponds to the RMS error over the entire retained image. All images are made from the same simulated VLA data of a 34 source field described in the text. }
	\label{fig:comparison}
\end{figure}

Table \ref{tab:rms-approx-error} lists the variants considered and the RMS approximation error computed over the entire image. The lines in Figure \ref{fig:comparison} show how the RMS value varies with $R$ for each of the methods, which are identified using the labels in the table. To reduce clutter in the figure, we have shown only the results for $x_0=y_0=0.25$. The example shown in Figure \ref{fig:Wstacking_FFT} corresponds to label (f). Because three-dimensional gridding is used, the cost of gridding each visibility is proportional to $W_xW_yW_z$.

For the original W-stacking algorithm, the recommended number of $w$-bins (or layers) given by equation (\ref{eq:N_w}) is $47$, corresponding to the case labelled (a). Two-dimensional least-misfit gridding is used with $W_x=W_y=7$, and $w$-binning effectively sets $W_z=1$ as usual. Although the gridding cost is one-seventh that of the case labelled (f), the RMS approximation error is more than $500000$ times worse. If we try increasing the number of $w$-layers to $470$ and then to $10000$ as in the cases labelled (b) and (c), the RMS error does decrease as expected, but the rate of decrease is rather slow as predicted by Figure~\ref{fig:improved_wstack_accuracy}. Even with $10000$ layers, which would incur a very large computational cost for performing the FFTs, the error is still about three orders of magnitude larger than for case (f).

If we perform three dimensional gridding, but use a convolutional function of smaller width, as in the cases labelled (d) and (e), we see that for $W=3$ in all three directions, the RMS error is already $27$ times smaller than that for the original W-stacking algorithm (case (a)). For case (d), the gridding cost and the number of layers (which sets the FFT cost) are both less for the improved algorithm. If we use $W=4$ in all three directions, the error is further reduced by at least another order of magnitude. Although the gridding cost of three-dimensional gridding with $W=4$ is $30\%$ more than for two-dimensional gridding with $W=7$, only $40\%$ of the $w$-layers are needed, reducing the cost of the FFT.

The cases labelled (h) and (i) show that one may also vary the image cropping parameters $x_0$ and $y_0$ in order to distribute the computational cost between gridding and the FFT. For $W=3$, the error can be reduced from that of case (d) by choosing $x_0=y_0=0.2$, but larger FFT sizes are needed because this decreases the portion of the image retained, increasing the cost of the FFT. On the other hand, with $W=4$, we can reduce the FFT cost compared to case (e) by choosing $x_0=y_0=0.33$, since this reduces both the number of layers and the number of points in the FFT, although at the cost of accuracy.

\begin{figure}
    \centering\includegraphics[width=\columnwidth]{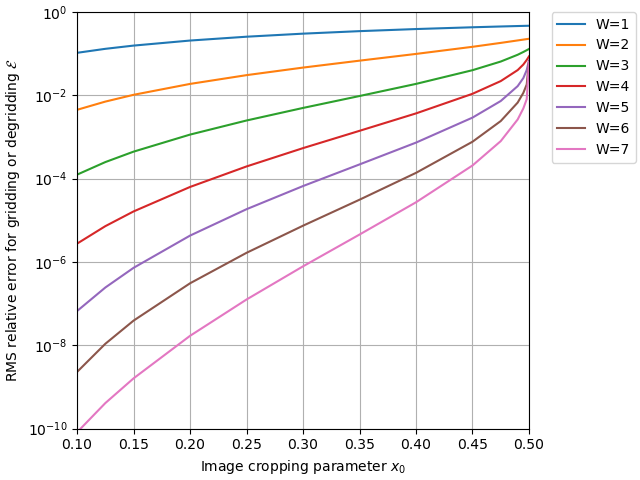}
    \caption{Dependence of the upper bound on the relative gridding and degridding errors in one dimension $\mathcal{E}$ on the choice of
    image cropping parameter ($x_0$ and $y_0$). The total error comprises contributions from each dimension in which gridding is performed. }
    \label{fig:accuracy_vs_image_cropping}
\end{figure}

%\color{green}
In order to quantify the effects of varying the image cropping parameters $x_0$ and $y_0$, Figure~\ref{fig:accuracy_vs_image_cropping} shows how the upper bound $\mathcal{E}$ on the gridding or degridding error in one dimension varies with these parameters and the width of the gridding convolution function when we use least-misfit gridding functions. Our recommended value of $0.25$ corresponds to discarding half the points calculated by the FFT in each map dimension, whereas a value of $0.5$ corresponds to using all the points. We see that reducing $x_0$ causes the errors to fall more rapidly with increasing $W$. This graph may be used to optimize further the trade off between the costs of the gridding and the fast Fourier transform, for a desired level of accuracy. In particular, the choice $x_0=y_0=\frac{1}{3}$ which corresponds to retaining $\frac{2}{3}$ of the points in each map dimension can make use of the efficient radix 3 FFT algorithm and requires approximately 40\% less transform time for the same sized map than with $x_0=y_0=0.25$ at the cost of only an order of magnitude degradation in accuracy for $W=7$.

\color{black}

In the theoretical development set out above, we learned that only the RMS error over the entire dirty image listed in Table~\ref{tab:rms-approx-error} is related to the error when using the adjoint operator for degridding and calculating visibilities from model data. It is nevertheless interesting to consider how the RMS approximation error varies over the different portions of the dirty image at differing distances from the map center. We see from Figure \ref{fig:comparison} that for the improved algorithm with $z$-offset (cases (d), (e) and (f)), the error varies by a factor of less than 3 over the whole image. For the original algorithm (cases labelled a, b and c), the error varies by many orders of magnitude, being small near the map center but increasing rapidly as one proceeds towards the edges. If, for instance, we consider lines (a) and (d), which correspond to the original algorithm with $47$ $w$-layers and the improved algorithm with $W=3$ and $18$ $w$-layers, we see that although the line for (a) lies below the line for (d) near the center, the area of this central region is about one twentieth of the whole image, and the error outside this region is so much larger that the error for case (a) is overall $27$ times larger than for case (d).

\begin{table}
\centering
 \caption{RMS approximation error of dirty image for various W-stacking methods. $W_x$, $W_y$ and $W_z$ are the number
 of points used for convolutional gridding in each direction, and $x_0$ and $y_0$ are the image cropping parameters.}
 \label{tab:rms-approx-error}
 \resizebox{\columnwidth}{!}{%
 \begin{tabular}{clcccccc}
  \hline
  Label & Algorithm & $W_x,W_y$ & $W_z$ & $W_x W_y W_z$ & $w$-layers & $x_0,y_0$ & RMS Error\\
  \hline
  a & Original & 7 & 1 & 49 & 47 & 0.25 & $9.8\times 10^{-3}$ \\
  b & Original & 7 & 1 & 49 & 470 & 0.25 & $6.4\times 10^{-4}$ \\
  c & Original & 7 & 1 & 49 & 10000 & 0.25 & $1.6\times 10^{-5}$ \\
  d & Improved & 3 & 3 & 27 & 18 & 0.25 & $3.6\times 10^{-4}$ \\
  e & Improved & 4 & 4 & 64 & 19 & 0.25 & $2.8\times 10^{-5}$ \\
  f & Improved & 7 & 7 & 343 & 22 & 0.25 & $1.8\times 10^{-8}$ \\
  g & Improved, no $z$ offset & 7 & 7 & 343 & 37 & 0.25 & $1.7\times 10^{-8}$ \\
  h & Improved & 3 & 3 & 27 & 22 & 0.20 & $1.6\times 10^{-4}$ \\
  i & Improved & 4 & 4 & 64 & 16 & 0.33 & $1.5\times 10^{-4}$ \\
  \hline
 \end{tabular}
 }
\end{table}

% Produeced by https://github.com/zoeye859/Wide-field-Nfacet
%W-stacking-improved Optimised with decreased w-planes (W=3, x0=0.2).ipynb

\begin{figure}
    a)\centering\includegraphics[width=0.8\columnwidth]{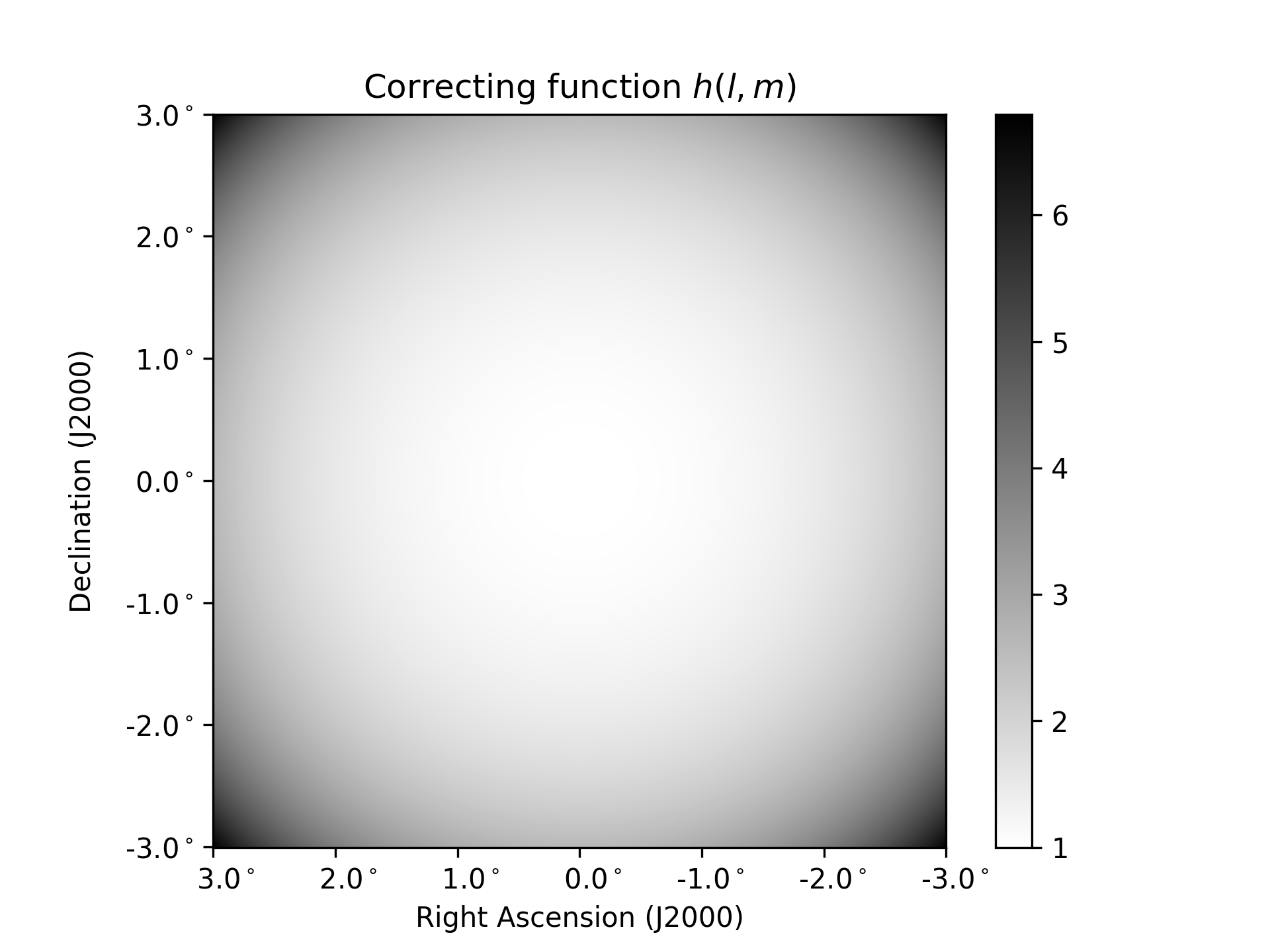}
    \hspace{3cm}
    b)\centering\includegraphics[width=0.8\columnwidth]{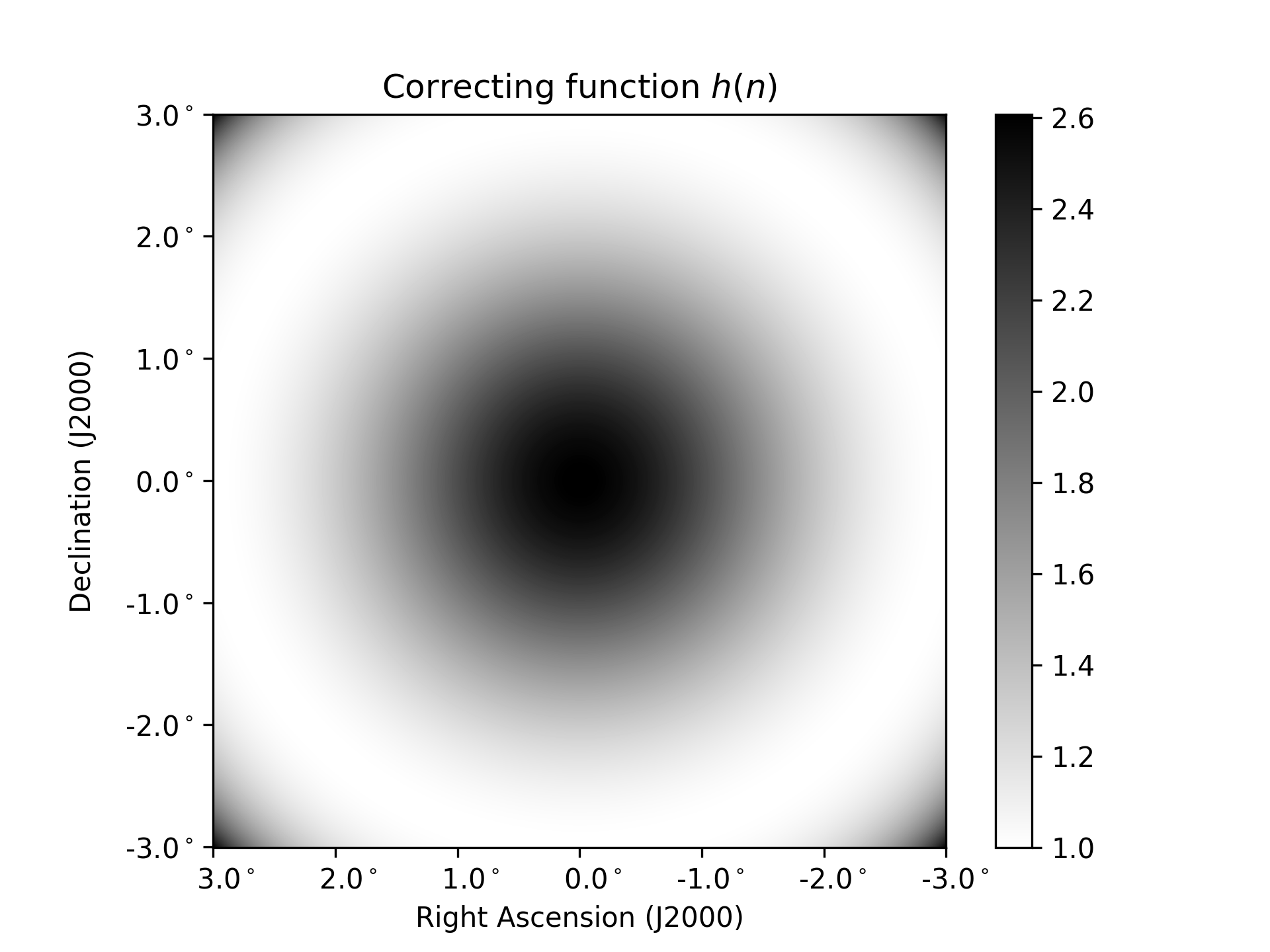}
    \caption {Factors of the gridding correction function for the improved W-stacking method with $W_x=W_y=W_z=7$ and $x_0=y_0=z_0=0.25$ (case (f) of Table~\ref{tab:rms-approx-error}). A $z$-offset has been included in order to minimize the number of $w$-layers needed. a) The correcting function for the $(l,m)$ plane, $h_x(l/[l_\mathrm{max}x_0])h_y(m/[m_\mathrm{max}y_0])$ b) The correcting function $h_z((n-n_0)/n_\mathrm{scale})$.}
    \label{fig:Wstacking_correction}
\end{figure}

For cases (d), (e) and (f) using the improved algorithm, the error rises slightly towards the center and edge of the image. This may be understood by considering Figure~\ref{fig:Wstacking_correction} which shows the gridding correction applied in calculating the image of Figure~\ref{fig:Wstacking_FFT}. As indicated by equation (\ref{eq:gridded-dirty-map}) and step 5 of Algorithm~5, the gridding correction is the product of three factors, $h_x h_y h_z$. In Figure~\ref{fig:Wstacking_correction}, the product $h_x h_y$ is shown in \ref{fig:Wstacking_correction}a and $h_z$ is shown in \ref{fig:Wstacking_correction}b. Since we are offsetting the $z=0$ plane so that it is not tangential to the celestial sphere at $n=1$ in order to reduce the number of $w$ layers, as described in the discussion following equation \ref{eq:N_{w'}+W}, $z=0$ occurs not at $x=y=0$ but in a ring of some finite radius in the image. Since $h_z$ is a minimum at $z=0$ and rises to a maximum at $z_0$, and the RMS approximation error of gridding tends to rise where the correction function is large, the minimum RMS approximation error occurs away from the image centre.

If we use the variant of the algorithm without $z$-offset as described by equation (\ref{eq:n_params_no_offset}), the minimum of the correction function can be placed at the map centre, at the cost of approximately doubling the number of $w$-layers. This has been done in the case labelled (g), for which the line in Figure~\ref{fig:comparison} does not rise as we approach the map centre. The overall approximation error shown in Table~\ref{tab:rms-approx-error} is only very slightly less than for case (f), however, so the additional computations associated with the extra layers may not be worthwhile.

Correct application of the full three-dimensional gridding correction function is essential to obtain accurate results using the improved $w$ stacking algorithm.

%\color{green}
Note that the entries in Table \ref{tab:rms-approx-error} are computed for a particular distribution of baselines and map size for which the product of non-planar terms $(n_\mathrm{max}-n_\mathrm{min})(w_\mathrm{max}-w_\mathrm{min})$ is such that a very large number of $w$ layers is required with the original W-stacking method before the 
gridding error in the $w$ direction becomes comparable with that in the $u$ and $v$ directions. Thus when comparing rows (a) and (d) for example, most of the RMS error in 
(a) arises from the $w$ direction. If the array is almost planar so that $w_\mathrm{max}-w_\mathrm{min}$ is very small, this is no longer the case, and at some point the 
RMS error for row (a) will drop below that for row (d), once the error is dominated by gridding in the $u$ and $v$ directions. In the limit of a planar array, the problem 
reduces to that of two-dimensional mapping, for which the advantage of larger values of $W_x$ and $W_y$ is a much higher level of accuracy, as discussed more fully in 
\citet{2020MNRAS.491.1146Y} and summarized in Figure~\ref{fig:accuracy_vs_image_cropping}. When used with Figure~\ref{fig:improved_wstack_accuracy}, quantitative estimates 
of the gridding errors in each of the three directions which contribute to the overall error may be made.
\color{black}

%W-stacking-improved Optimised with decreased w-planes (images produced for paper).ipynb

\begin{figure}
    \centering
    \includegraphics[width=\columnwidth]{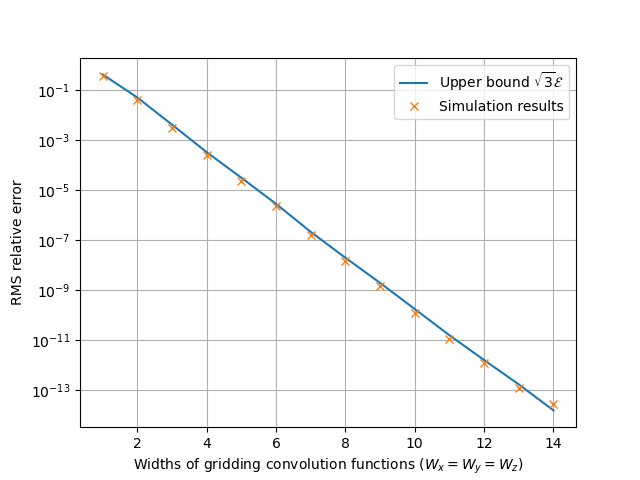}
    \caption{Variation of accuracy of gridding or degridding algorithms with the width of support of the convolving functions ($W_x=W_y=W_z$) for image cropping parameters $x_0=y_0=z_0=0.25$.
    The line shows the upper bound $\sqrt{3}\mathcal{E}$ on the approximation error of three-dimensional gridding or degridding for least-misfit optimal functions (equation \ref{eq:norm_bound}). The crosses are the results of computing the RMS relative error between the exact dirty map calculated using the DFT and the approximation calculated using FFT and three dimensional convolutional gridding with identical least-misfit optimal functions in all directions, using the improved W-stacking algorithm with simulated data, as in Figure \ref{fig:Wstacking_FFT}.
    }\label{fig:rms_relative_error}
\end{figure}

We next present the results of numerical simulations again using the synthetic VLA data and the improved W-stacking method with the same least-misfit convolutional gridding functions in all three directions. The aim is to show how well the one-dimensional upper bound $\mathcal{E}$ given by equation (\ref{eq:norm_bound}) constrains the RMS approximation error of the dirty images. We compute $\mathcal{E}$ for $x_0=0.25$, and values of $W$ from $1$ and $14$. Since this applies to one-dimensional gridding in each of the three directions, and the errors are approximately equal in all directions when the same gridding function is used, we multiply the bound by $\sqrt{3}$. In Figure~\ref{fig:rms_relative_error}, the line is drawn through the computed value of $\sqrt{3}\mathcal{E}$, and the crosses show the RMS approximation error calculated from the simulations, normalized by the RMS of the visibilities as specified by the quotient on the right-hand side of equation (\ref{eq:norm_bound2}) without the $4l_\mathrm{max}m_\mathrm{max}$ factor (see the discussion preceding equation (\ref{eq:norm_bound})). Double precision arithmetic has been used in the simulations, and we see that all lie slightly below the upper bounds, as expected, except for the last point $W=14$, which has reached the limit of the precision available. For single precision arithmetic, full accuracy is achieved for $W=7$ and $x_0=0.25$. This figure is also useful in estimating the accuracy for differing choices of $W$ in the improved W-stacking method, to control the computation costs for specific data sets.

\section{Application to real data}\label{sec:application}

This section applies the improved W-stacking method to two sets of real data. As shown in Section \ref{sec:error_bounds}, a consequence of the adjoint relationship between the dirty image making and degridding is that bounds (appropriately defined) on the accuracy with which each can be performed are essentially identical. Therefore, although we use dirty images in this section to illustrate various features, the accuracy of these dirty images also indicates that visibilities calculated from models using the degridding process are correspondingly accurate. As a final example, we present a reconstructed image of the whole sky from synthetic data, using the Maximum Entropy method, which employs both the gridding and degridding forms of the improved W-stacking method.

The first real dataset comprises a VLA D-array observation of the supernova remnant G55.7+3.4 \footnote{Data available at \url{https://casaguides.nrao.edu/index.php/EVLA_Wide-Band_Wide-Field_Imaging:_G55.7_3.4-CASA4.4}}.

\begin{figure}
    a)\centering\includegraphics[width=1.1\columnwidth]{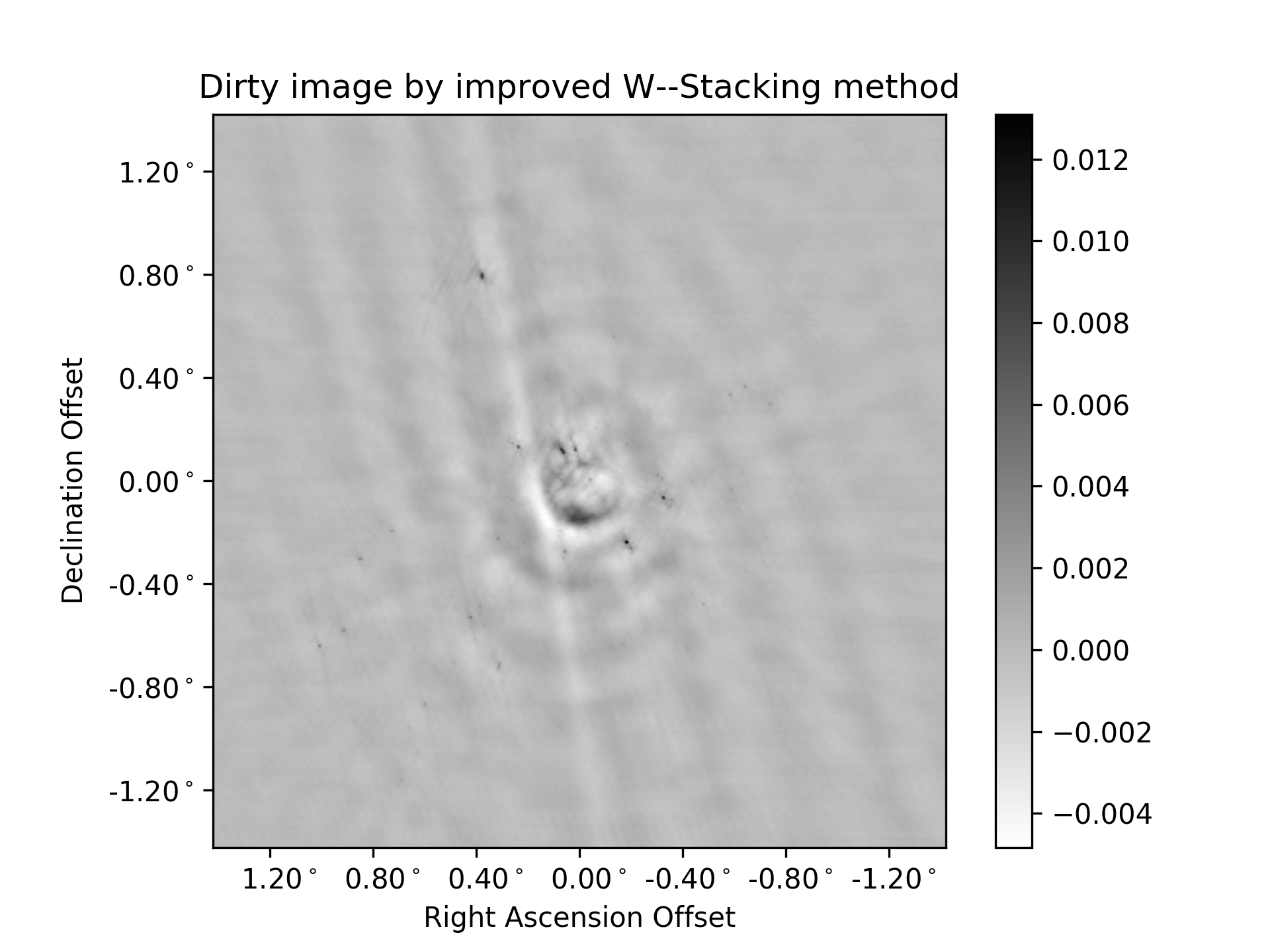}
    \hfill
    b)\centering\includegraphics[width=1.1\columnwidth]{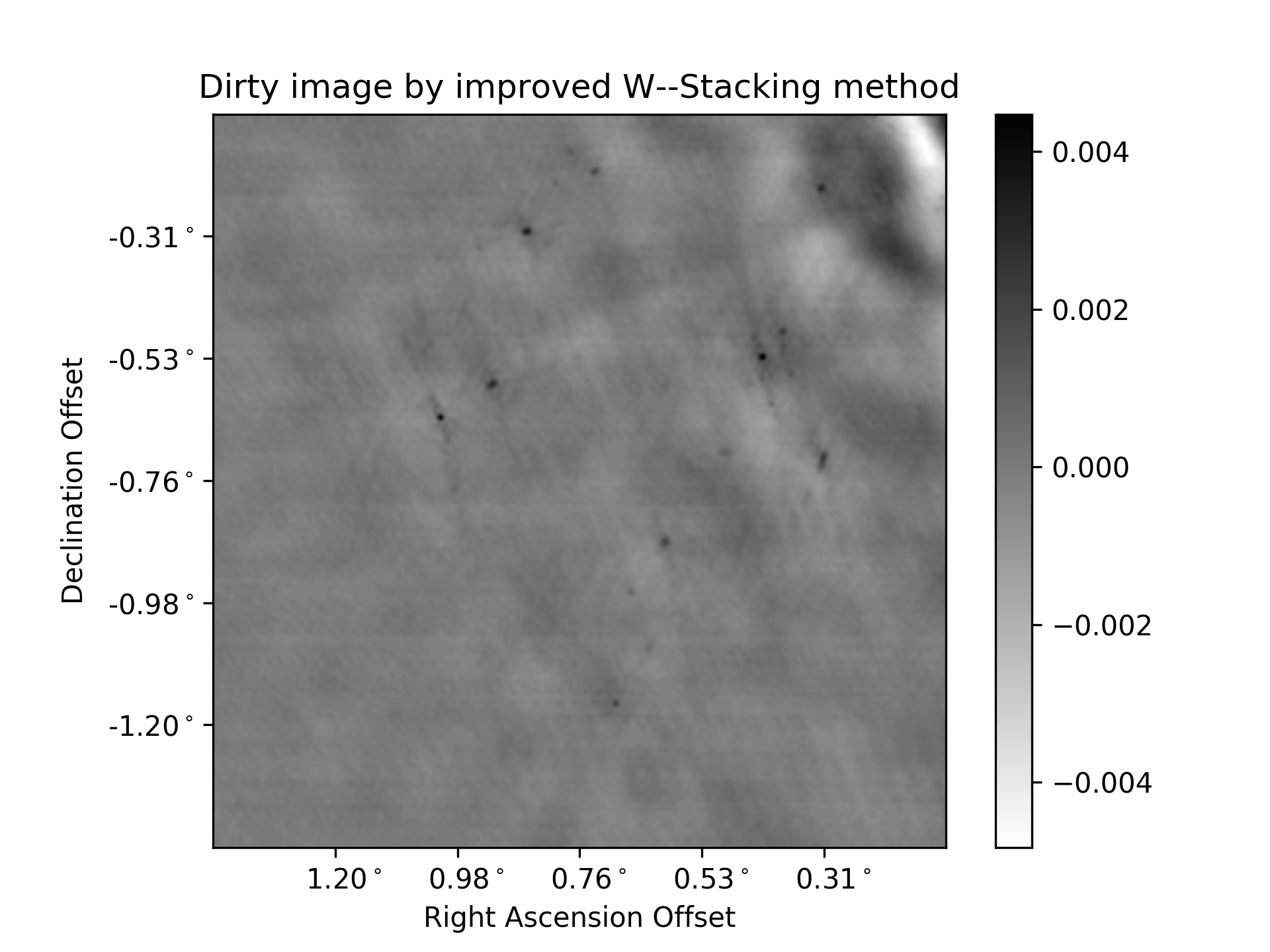}
    \caption{a) Celestial sphere dirty image of the supernova remnant G55.7+3.4, generated using the improved W-stacking method. The field of view is centred on $(19^h21^m40^s,+21^\circ 45')$. The ripples on the image are due to the use of natural weighting. b) Zoomed-in window of the bottom left portion of the dirty image. The two images are shown with the $x$-axis representing the right ascension offset and the $y$-axis representing the declination offset.}
    \label{fig:wide_real_dirtymap_improve}
\end{figure}
%/appch/data1/hy297/CASA-tutorial-wide/Results from Steve.ipynb

We applied the improved W-stacking method to the calibrated data. All four spectral window data were imaged. For each image, the fluxes have been adjusted using a spectral index of $0.5$. The resulting averaged celestial sphere dirty image is shown in Figure \ref{fig:wide_real_dirtymap_improve}. Ripples in the image are due to the use of natural weighting. The image size is 2560 by 2560, with a pixel size of $4$ arcsec. The field of view is centred on $(19^h21^m40^s,+21^\circ 45')$. The image is shown with the right ascension offset as the $x$-axis and the declination offset as the $y$-axis. The image at the bottom is an enlargement of the lower left corner of the top image. The improved W-stacking method successfully removes the prominent arcs around the point sources far from the phase centre, which would be present if no wide-field imaging methods were used.

\begin{figure}
    a)\centering\includegraphics[width=\columnwidth]{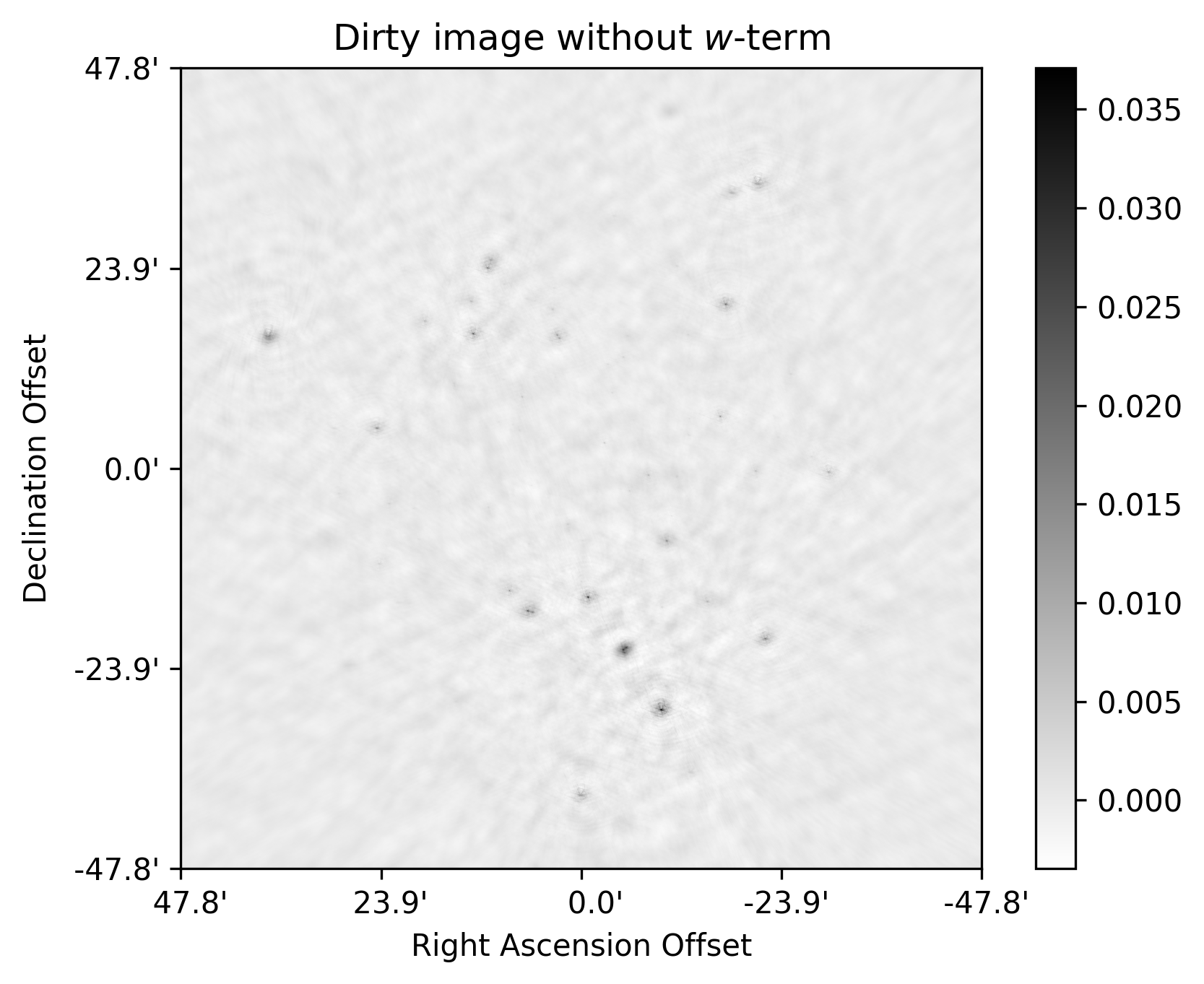}
    \hfill
    b)\centering\includegraphics[width=\columnwidth]{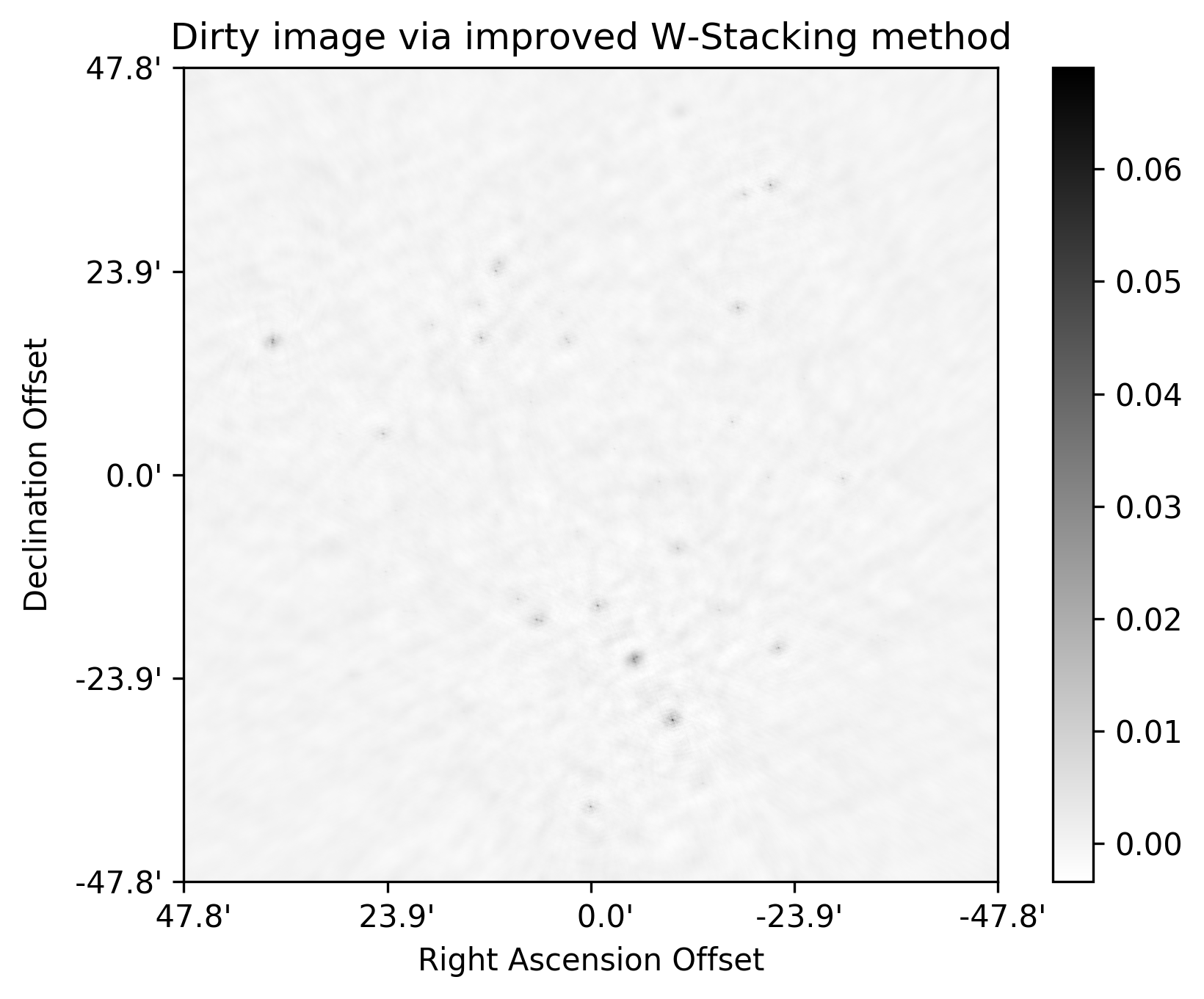}
    \caption{a) Tangent dirty image of GMRT data. b) Celestial sphere dirty image made using the improved W-stacking method. The field of view is centred on $(00^h23^m10^s, +31^\circ 53')$. The $x$-axis represents the right ascension offset, and the $y$-axis the declination offset.}
    \label{fig:wide_wstacking_dirtymap}
\end{figure}

The second dataset comprises GMRT data observed at frequencies ranging from 32 MHz to 610 MHz in 24 channels. This data was presented in \citet{2017MNRAS.464.3357W}, and its wide-field imaging was performed in \texttt{AIPS} using multiple image facets. The observations cover the field AMI001 of 0.84 $\mathrm{deg}^2$ centred on $(00^h23^m10^s, +31^\circ 53')$. We apply the improved W-stacking method to the data. The tangent plane dirty image and the celestial sphere dirty image are shown in Figure \ref{fig:wide_wstacking_dirtymap}, at top and bottom. Both images are $4096$ by $4096$ in size, with a cell size of $1.4$ arcsec. The field of view is centred on $(00^h23^m10^s, +31^\circ 53')$. The two images are shown with the right ascension offset as the $x$-axis and the declination offset as the $y$-axis.

\begin{figure}
    a)\centering\includegraphics[width=\columnwidth]{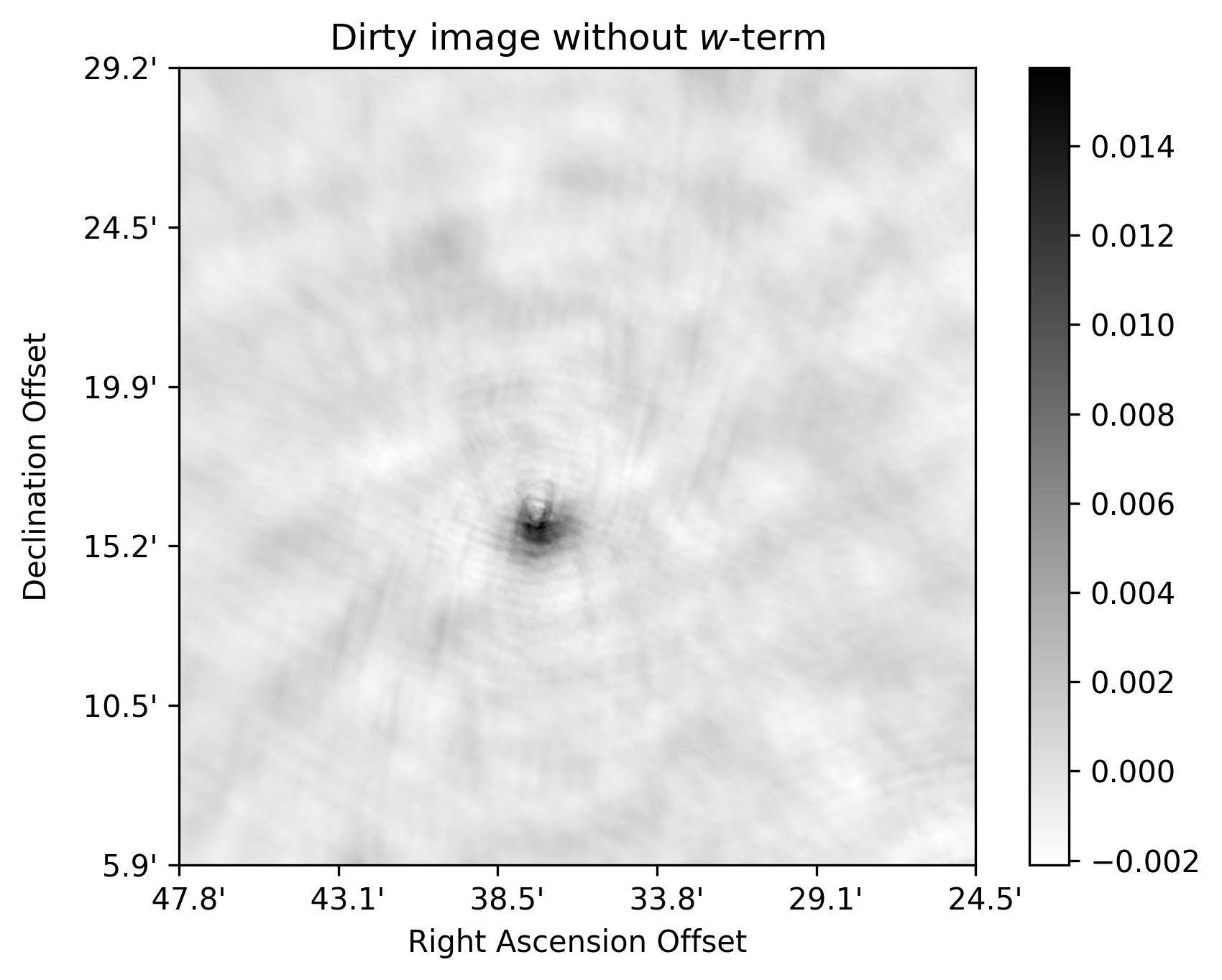}
    b)\centering\includegraphics[width=\columnwidth]{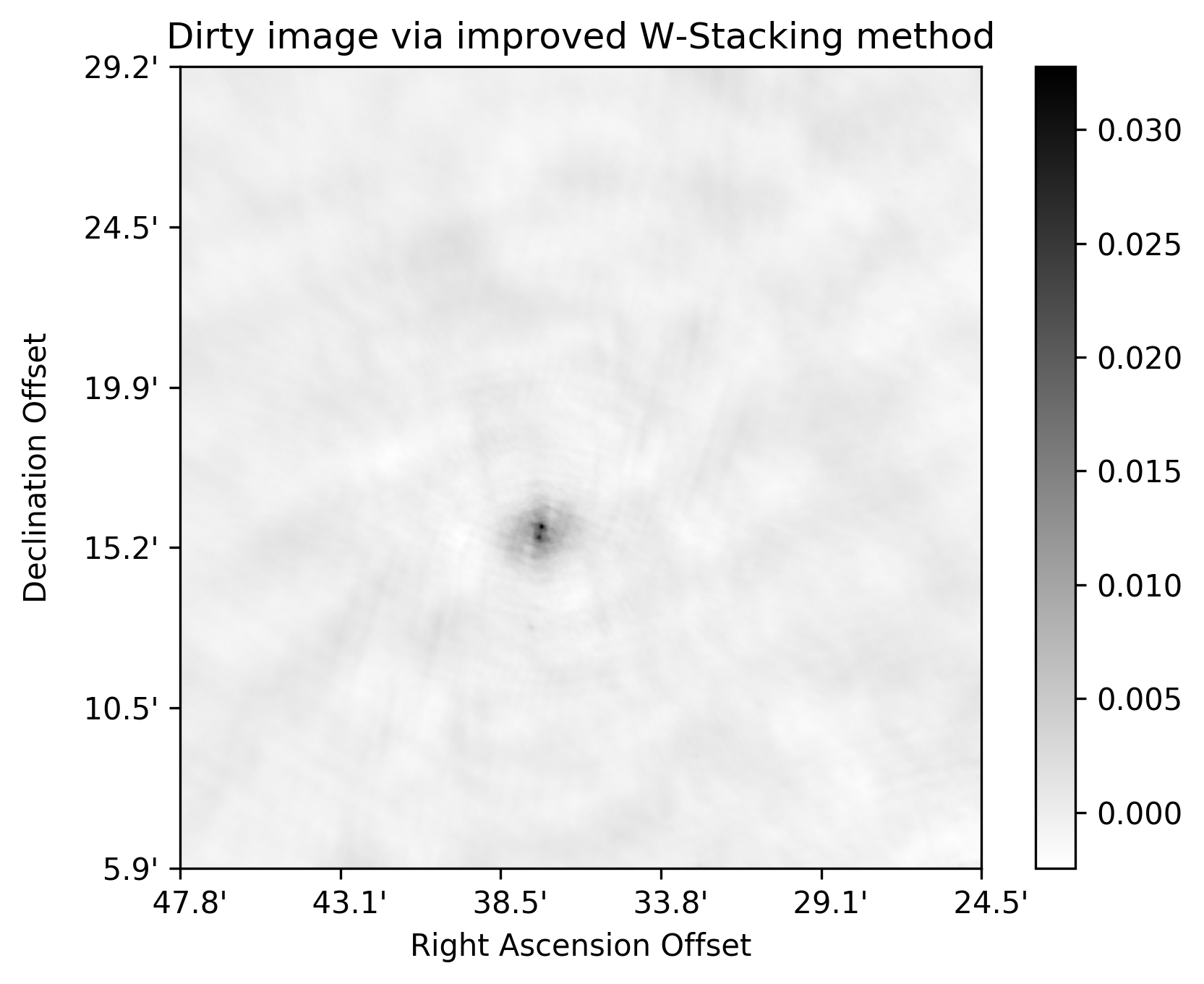}
    \caption{Two nearby sources with considerable offsets from the phase centre $(00^h23^m10^s, +31^\circ 53')$ are shown with prominent arcs in the enlarged tangent plane dirty image, at top. The improved W-stacking method removes the non-coplanar effects, as shown on the lower map; the two sources are resolved.}
    \label{fig:wide_wstacking_dirtymap_zoom}
\end{figure}

In Figure \ref{fig:wide_wstacking_dirtymap_zoom} we zoom in on the top left portion of both images. In the tangent plane dirty image displayed at the top, prominent arcs are visible around the two nearby point sources, which are far from the phase centre. These artefacts make it impossible to confirm the existence of the second point source. Use of the improved W--stacking method removes the artefacts and allows the two point sources to be distinguished (as in the bottom map).

We made the dirty image using the improved W-stacking method on a computer with a 10-core Intel i9-9820X processor and 16GB memory. The least-misfit gridding function ($W=7$ and $x_0=0.25$) was used with a pre-calculated look-up table of $10^6$ points per pixel, and $1.2$ seconds were taken to calculate the gridding function values for all visibility data ($415,613$ entries). The gridding process then took 19 seconds. A total of $40$ $w$ layers were used. The FFT operation took the longest clock time, 304 seconds, with an image size of $4096$ by $4096$. By taking advantage of the multiple cores, the FFT time was successfully reduced to 114 seconds. The image was made without $z$-offset, making $z=0$ coincide with the map centre. If we do use a $z$ offset in the optimal way, equation \ref{eq:N_{w'}+W} shows that the number of $w$ layers can be further reduced to $24$, with a corresponding reduction in the computation time.

To demonstrate the use of the improved W-stacking method in wide-field image reconstruction, we present an example using the Maximum Entropy method package
MemSys5 \citep{gull1991quantified}. The parameters have been chosen so that the point spread function varies significantly over the field, so that reconstruction is not simply a deconvolution. The package requires two routines to be written; one generates the visibilities from a given image model, and the second calculates the transpose (or adjoint) of this operation. The latter operation is performed by the improved W-stacking mapper using the least-misfit gridding functions. This routine is called ``Tropus'' in MemSys terminology\footnote{Geoff Daniell named this before the publication of \citet{1978Natur.272..686G}}.
The former operation is performed using the associated degridding algorithm outlined above, and is called ``Opus''.
Within the Memsys5 package, the subroutine ``MemTrq'' is provided to test the consistency of the Opus/Tropus pair, and this test was passed at a level of a few times $10^{-10}$ for the present application.

\begin{figure}
a)\centering\includegraphics[width=\columnwidth]{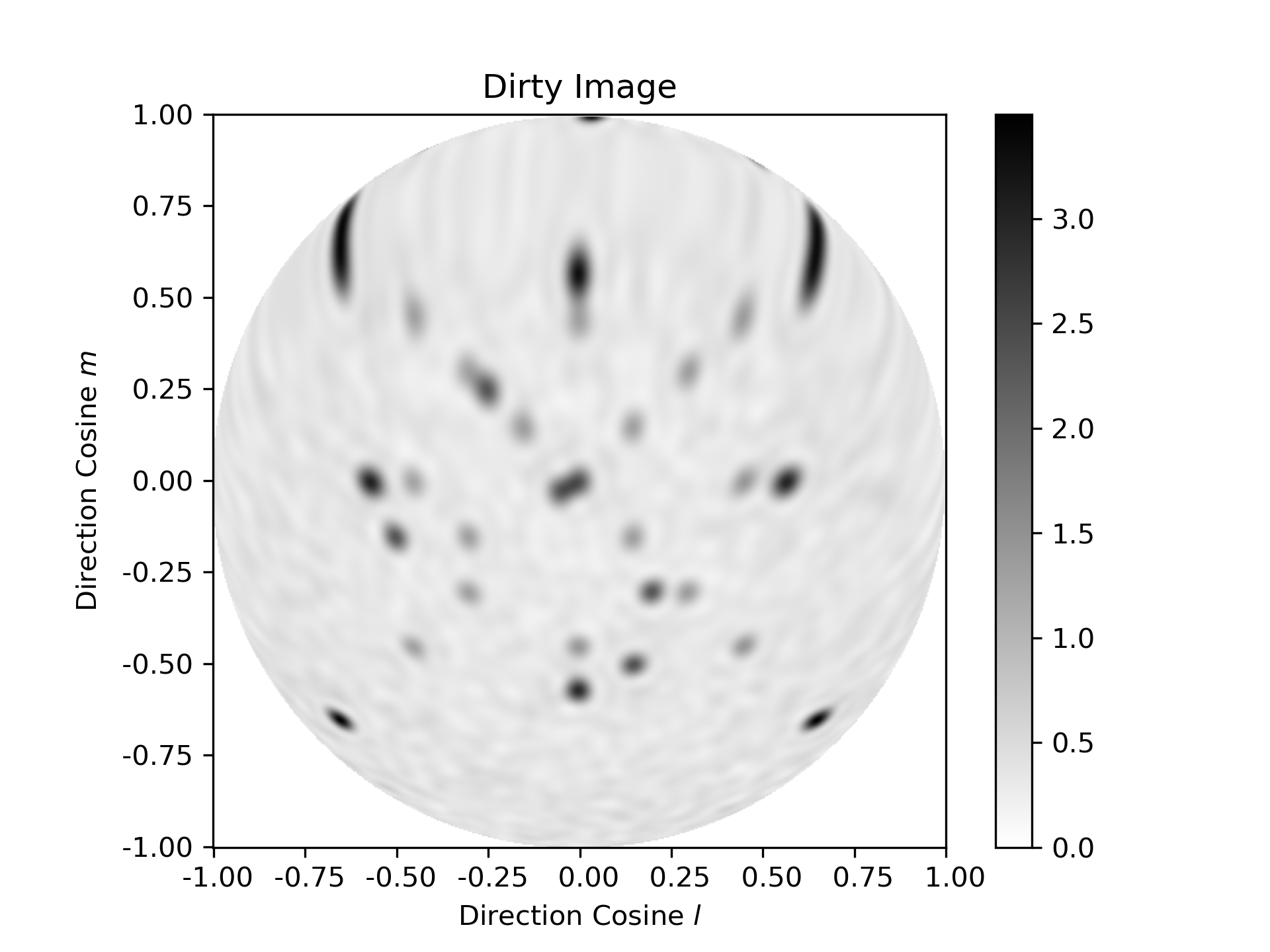}
b)\centering\includegraphics[width=\columnwidth]{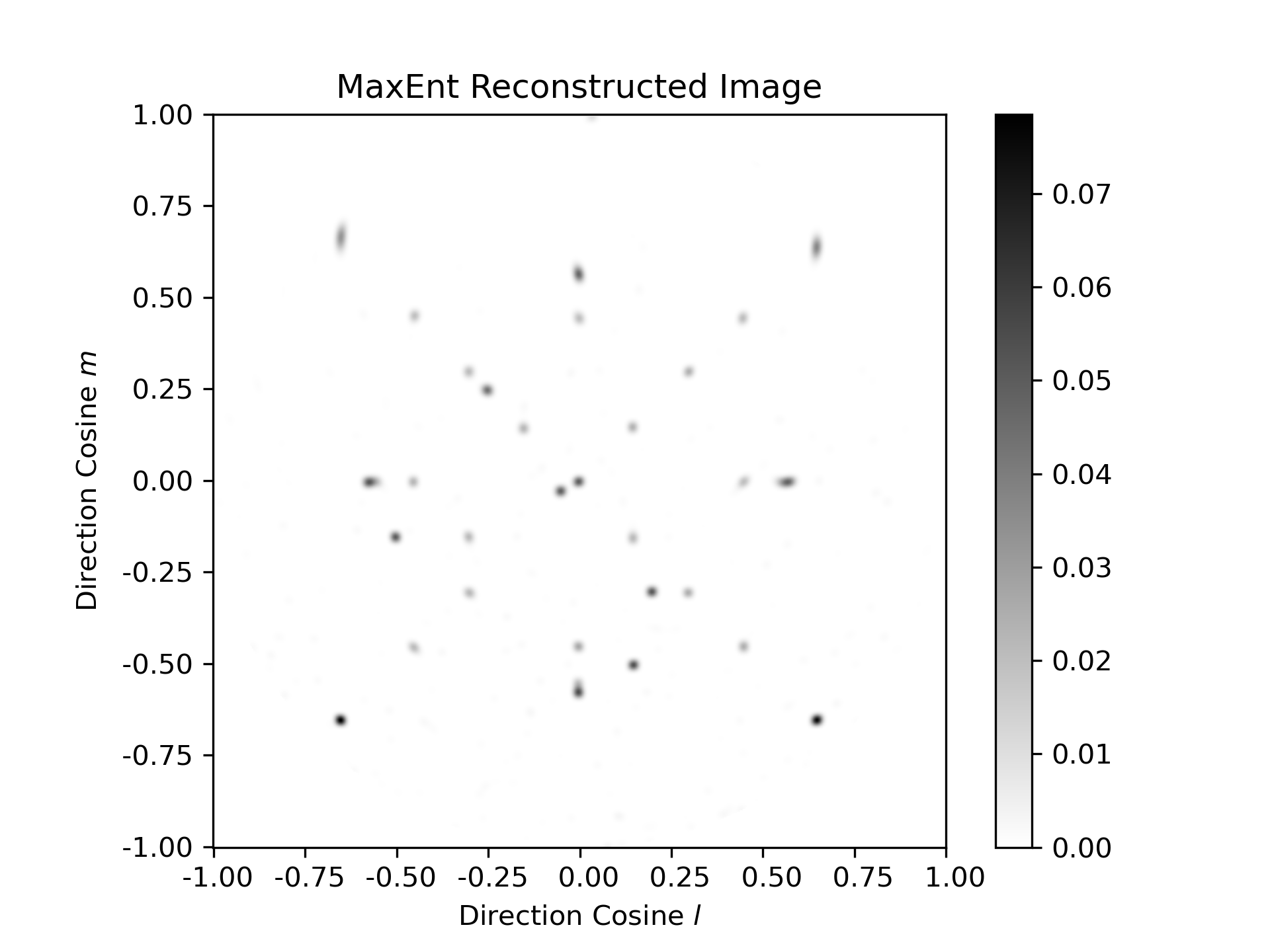}
c)\centering\includegraphics[width=\columnwidth]{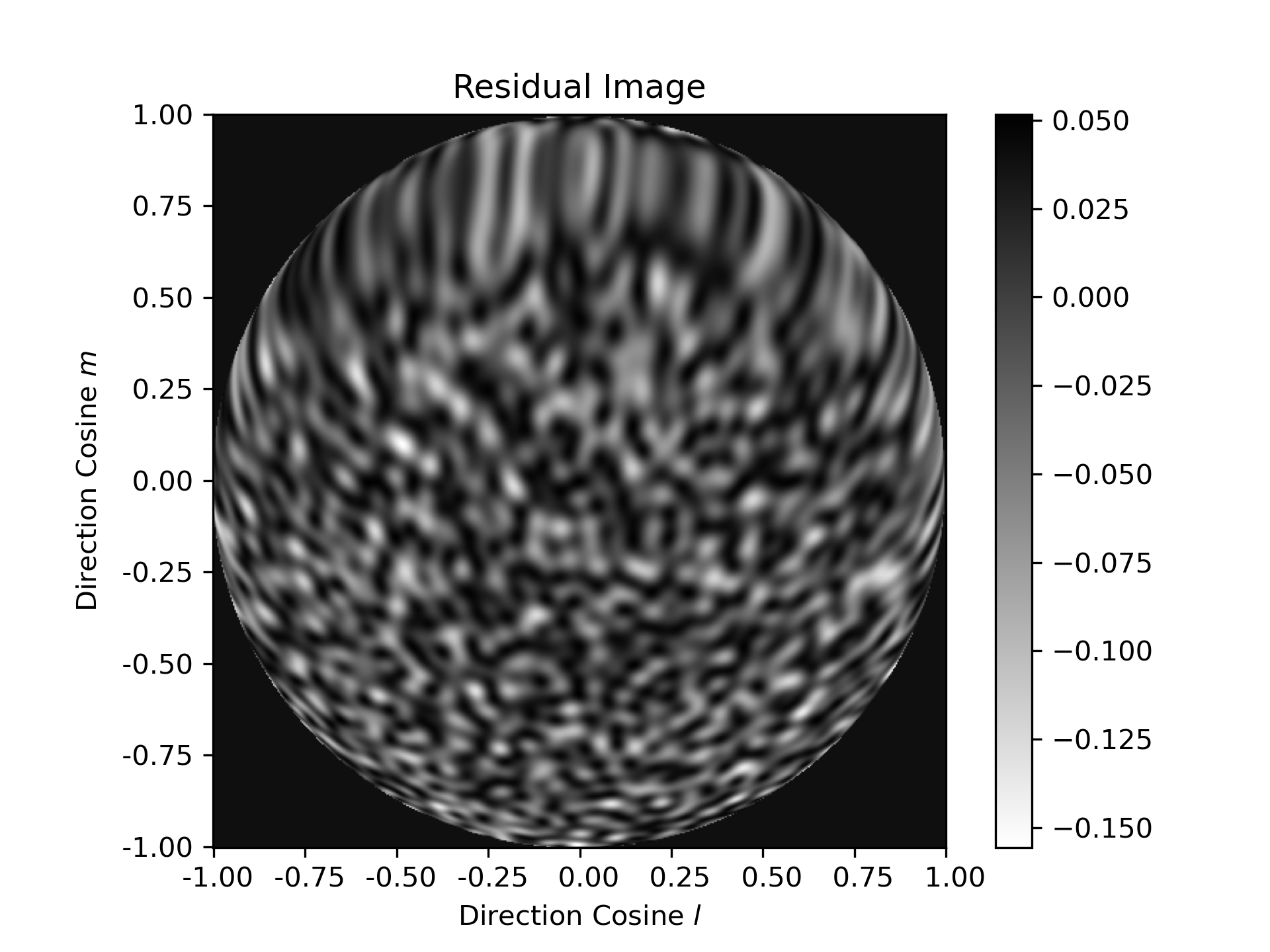}
\caption{(a) Dirty image made from the VLA A-array configuration with 32,896 baselines. The ranges of $(l,m)$ are $\pm 1$. There are 34 sources with a maximum flux of 3~Jy, and a Gaussian noise of 8~Jy been added to the real and imaginary parts of all visibilities, such that the noise level on the dirty image is expected to be 0.044~Jy. The output image size is $512\times 512$ pixels. The maps shown here are the Cartesian projection of the whole Celestial hemisphere. The centre of the map is at Declination 90$^\circ$.
The outer circle has Declination 0$^\circ$ and represents 24~hours of Right Ascension.
(b) MaxEnt reconstruction from this Dirty image. For display purposes, the output was convolved with a Gaussian of width 2.24 pixels.
(c) Dirty residuals made from this reconstruction, which appear to be entirely noise.}
    \label{fig:wmem8}
\end{figure}

The simulated data using the VLA A-array configuration\footnote{For illustrative purposes, the $uvw$ values were rescaled so as to reduce the resolution.} contains 34 point sources, whose fluxes and positions are given in the Appendix. The maximum source flux is 3~Jy. A Gaussian noise of 8~Jy has been added to the real and imaginary parts of all visibilities, which implies that the noise level on the dirty image should be 0.044~Jy. The ranges of $(l,m)$ are $\pm 1$. The output image size is $512\times 512$ pixels. Figure~\ref{fig:wmem8} shows the dirty image made by the improved W-stacking method, shows the reconstructed image using the MemSys5, and shows the corresponding residuals, which appear to be entirely noise. These maps are the Cartesian projection of the whole Celestial hemisphere. The centre of the map is at Declination 90$^\circ$. The outer circle has Declination 0$^\circ$ and represents 24~hours of Right Ascension.

The units of the MaxEnt map are Janskys per pixel, and must be integrated over areas
in order to determine the total flux. To this end, and to improve the display, the MaxEnt reconstruction has been convolved with a PSF of 2.24 pixels, corresponding to MemSys's
ICF of 4\footnote{The MemSys5 manual is available online at \url{http://www.mrao.cam.ac.uk/~steve/MemSys5_manual.pdf} and can be consulted for details.}.

In summary, the first two experiments confirm that the improved W-stacking method works well with real observational data, and the third experiment demonstrates that deconvolution methods such as the Maximum Entropy method can be used to deconvolve the sky brightness from the dirty image obtained from the improved W-stacking method. %For observations with a large $w$ range, the improved W--stacking method is recommended in order to obtain an image with higher dynamic range and potentially lower computational cost.

\section{Conclusions}\label{sec:conclusion}

We began this paper by introducing the wide-field imaging problem involving the complex $w$-term. The $w$-term cannot be neglected when the Fresnel number $N_F<1$, and a wide-field imaging method is needed to relate the sky brightness and visibilities accurately. %Even with planar arrays, large values of $w$ can still exist if the observation is not normal to the plane of the array. Experiments have shown that for a planar array whose values of $w$ and $v$ are almost perfectly correlated (in the linear relation $w=\alpha v$), the dirty image of the celestial sphere merely undergoes a displacement from the dirty image of the tangent plane by an amount $-\alpha n$ along the declination axis.
After introducing the original W-stacking method, we proposed our improved W-stacking method, which minimizes the degridding and dirty image approximation error relative to the exact results given by direct computation of the Fourier transform. This involves the use of three-dimensional gridding convolution functions, gridding correction and image cropping. By using the least-misfit gridding convolution and correction functions in each dimension, the accuracy and number of $w$-layers can be controlled by choosing the width $W$ of the convolution function and the image cropping parameter $x_0$.

By applying the improved W-stacking method, the misfit between the DFT and FFT dirty images on the celestial sphere is reduced to the limit of single precision arithmetic when $W_x=W_y=W_z=7$
and $x_0=y_0=z_0=0.25$. Similarly, the limit of double precision arithmetic can be reached with $W_x=W_y=W_z=14$ and $x_0=y_0=z_0=0.25$.

To achieve single precision accuracy using the original W-stacking method, in contrast, the number of $w$ layers required can exceed those required for improved W-stacking by a factor of at least six orders of magnitude. The additional cost associated with three-dimensional gridding in the improved method is more than outweighed by the increased FFT costs required by the original to attain the same level of accuracy.

If less accuracy is adequate, accuracy comparable to the original W-stacking method can be attained by reducing the width of the convolution functions in all three directions, so that the gridding and FFT costs can both be reduced. The approximation error using the improved method is distributed more evenly over the image than with the original method, for which the error is typically small only very close to the map centre.

The original W-stacking method may be analyzed as a special case of three-dimensional gridding in which $W_z=1$ and the gridding convolution function on $w$-axis is a top hat. The error introduced due to a small $W_z$ dominates that in the $x$ and $y$ directions for most choices of parameters in which the non-planar terms are important, even though $W_x=W_y=7$ are used in the current practice of the original W-stacking method in \texttt{WSCLEAN}. With the improved W-stacking method, it becomes feasible to divide the error budget evenly between the gridding processes in all three dimensions, even when mapping large areas of the sky, and to balance the cost of gridding and computing FFTs as necessary.

\color{black}
\section*{Acknowledgements}

%We thank the anonymous referee for his/her comments.

We are grateful for Andr\'e Offringa's comments and a suggestion which led to a reduction in the number of $w$ layers required in the $w$-direction. We also thank Martin Reinecke and Philipp Arras for their initiative and efforts to implement our algorithm into software \texttt{NIFTY} and \texttt{WSCLEAN}, and
we are grateful to Anton Garrett for his
careful editorial advice.

\section*{Data Availability}
Data available on request.
%%%%%%%%%%%%%%%%%%%%%%%%%%%%%%%%%%%%%%%%%%%%%%%%%%



\bibliographystyle{mnras}
\bibliography{reference.bib}

\begin{thebibliography}{}
\makeatletter
\relax
\def\mn@urlcharsother{\let\do\@makeother \do\$\do\&\do\#\do\^\do\_\do\%\do\~}
\def\mn@doi{\begingroup\mn@urlcharsother \@ifnextchar [ {\mn@doi@}
  {\mn@doi@[]}}
\def\mn@doi@[#1]#2{\def\@tempa{#1}\ifx\@tempa\@empty \href
  {http://dx.doi.org/#2} {doi:#2}\else \href {http://dx.doi.org/#2} {#1}\fi
  \endgroup}
\def\mn@eprint#1#2{\mn@eprint@#1:#2::\@nil}
\def\mn@eprint@arXiv#1{\href {http://arxiv.org/abs/#1} {{\tt arXiv:#1}}}
\def\mn@eprint@dblp#1{\href {http://dblp.uni-trier.de/rec/bibtex/#1.xml}
  {dblp:#1}}
\def\mn@eprint@#1:#2:#3:#4\@nil{\def\@tempa {#1}\def\@tempb {#2}\def\@tempc
  {#3}\ifx \@tempc \@empty \let \@tempc \@tempb \let \@tempb \@tempa \fi \ifx
  \@tempb \@empty \def\@tempb {arXiv}\fi \@ifundefined
  {mn@eprint@\@tempb}{\@tempb:\@tempc}{\expandafter \expandafter \csname
  mn@eprint@\@tempb\endcsname \expandafter{\@tempc}}}

\bibitem[\protect\citeauthoryear{{Adams} et~al.,}{{Adams}
  et~al.}{2020}]{2020AAS...23513607A}
{Adams} E.~A.,  et~al., 2020, in American Astronomical Society Meeting
  Abstracts \#235. p. 136.07

\bibitem[\protect\citeauthoryear{{Arras}, {Reinecke}, {Westermann}  \&
  {En{\ss}lin}}{{Arras} et~al.}{2020}]{2020arXiv201010122A}
{Arras} P.,  {Reinecke} M.,  {Westermann} R.,   {En{\ss}lin} T.~A.,  2020,
  arXiv e-prints, \href {https://ui.adsabs.harvard.edu/abs/2020arXiv201010122A}
  {p. arXiv:2010.10122}

\bibitem[\protect\citeauthoryear{{Bhatnagar}, {Rau}  \& {Golap}}{{Bhatnagar}
  et~al.}{2013}]{2013ApJ...770...91B}
{Bhatnagar} S.,  {Rau} U.,   {Golap} K.,  2013, \mn@doi [\apj]
  {10.1088/0004-637X/770/2/91}, \href
  {https://ui.adsabs.harvard.edu/abs/2013ApJ...770...91B} {770, 91}

\bibitem[\protect\citeauthoryear{{Clark}}{{Clark}}{1980}]{1980A&A....89..377C}
{Clark} B.~G.,  1980, \aap, \href
  {https://ui.adsabs.harvard.edu/abs/1980A&A....89..377C} {89, 377}

\bibitem[\protect\citeauthoryear{Conway}{Conway}{1990}]{Conway1990_FunctionalAnalysis}
Conway J.~B.,  1990, A Course in Functional Analysis, Second Edition.
Springer-Verlag, New York

\bibitem[\protect\citeauthoryear{Cooley \& Tukey}{Cooley \&
  Tukey}{1965}]{cooley1965algorithm}
Cooley J.~W.,  Tukey J.~W.,  1965, Mathematics of Computation, 19, 297

\bibitem[\protect\citeauthoryear{{Cornwell}}{{Cornwell}}{2008}]{2008ISTSP...2..793C}
{Cornwell} T.~J.,  2008, \mn@doi [IEEE Journal of Selected Topics in Signal
  Processing] {10.1109/JSTSP.2008.2006388}, \href
  {https://ui.adsabs.harvard.edu/abs/2008ISTSP...2..793C} {2, 793}

\bibitem[\protect\citeauthoryear{{Cornwell}, {Golap}  \&
  {Bhatnagar}}{{Cornwell} et~al.}{2003}]{Cornwell2003W}
{Cornwell} T.~J.,  {Golap} K.,   {Bhatnagar} S.,  2003, Technical report, EVLA
  Memo 67

\bibitem[\protect\citeauthoryear{{Cornwell}, {Golap}  \&
  {Bhatnagar}}{{Cornwell} et~al.}{2005}]{2005ASPC..347...86C}
{Cornwell} T.~J.,  {Golap} K.,   {Bhatnagar} S.,  2005, in {Shopbell} P.,
  {Britton} M.,   {Ebert} R.,  eds,  Astronomical Society of the Pacific
  Conference Series Vol. 347, Astronomical Data Analysis Software and Systems
  XIV. p.~86

\bibitem[\protect\citeauthoryear{{Cornwell}, {Voronkov}  \&
  {Humphreys}}{{Cornwell} et~al.}{2012}]{2012SPIE.8500E..0LC}
{Cornwell} T.~J.,  {Voronkov} M.~A.,   {Humphreys} B.,  2012, in {Bones} P.~J.,
   {Fiddy} M.~A.,   {Millane} R.~P.,  eds,  Society of Photo-Optical
  Instrumentation Engineers (SPIE) Conference Series Vol. 8500, Image
  Reconstruction from Incomplete Data VII. p. 85000L (\mn@eprint {arXiv}
  {1207.5861}), \mn@doi{10.1117/12.929336}

\bibitem[\protect\citeauthoryear{Gull}{Gull}{1989}]{gull1989developments}
Gull S.~F.,  1989, Developments in Maximum Entropy Data Analysis.
Springer Netherlands, Dordrecht, pp 53--71,
  \mn@doi{10.1007/978-94-015-7860-8_4}, \url
  {https://doi.org/10.1007/978-94-015-7860-8_4}

\bibitem[\protect\citeauthoryear{{Gull} \& {Daniell}}{{Gull} \&
  {Daniell}}{1978}]{1978Natur.272..686G}
{Gull} S.~F.,  {Daniell} G.~J.,  1978, \mn@doi [\nat] {10.1038/272686a0}, \href
  {https://ui.adsabs.harvard.edu/abs/1978Natur.272..686G} {272, 686}

\bibitem[\protect\citeauthoryear{Gull \& Skilling}{Gull \&
  Skilling}{1984}]{gull1984maximum}
Gull S.~F.,  Skilling J.,  1984, \mn@doi [IEE Proceedings F - Communications,
  Radar and Signal Processing] {10.1049/ip-f-1:19840099}, 131, 646

\bibitem[\protect\citeauthoryear{Gull \& Skilling}{Gull \&
  Skilling}{1991}]{gull1991quantified}
Gull S.,  Skilling J.,  1991, Quantified Maximum Entropy MemSys5 User's Manual.
Maximum Entropy Data Consultants, \url
  {http://www.mrao.cam.ac.uk/~steve/malta2009/images/MemSys5_manual.pdf}

\bibitem[\protect\citeauthoryear{{Hoang} et~al.,}{{Hoang}
  et~al.}{2018}]{2018MNRAS.478.2218H}
{Hoang} D.~N.,  et~al., 2018, \mn@doi [\mnras] {10.1093/mnras/sty1123}, \href
  {https://ui.adsabs.harvard.edu/abs/2018MNRAS.478.2218H} {478, 2218}

\bibitem[\protect\citeauthoryear{{H{\"o}gbom}}{{H{\"o}gbom}}{1974}]{1974A&AS...15..417H}
{H{\"o}gbom} J.~A.,  1974, \aaps, \href
  {https://ui.adsabs.harvard.edu/abs/1974A&AS...15..417H} {15, 417}

\bibitem[\protect\citeauthoryear{Humphreys \& Cornwell}{Humphreys \&
  Cornwell}{2011}]{humphreys2011analysis}
Humphreys B.,  Cornwell T.,  2011, Square Kilometre Array Memo No. 132, 132

\bibitem[\protect\citeauthoryear{{Intema}, {Jagannathan}, {Mooley}  \&
  {Frail}}{{Intema} et~al.}{2017}]{2017A&A...598A..78I}
{Intema} H.~T.,  {Jagannathan} P.,  {Mooley} K.~P.,   {Frail} D.~A.,  2017,
  \mn@doi [\aap] {10.1051/0004-6361/201628536}, \href
  {https://ui.adsabs.harvard.edu/abs/2017A&A...598A..78I} {598, A78}

\bibitem[\protect\citeauthoryear{{Kapinska}}{{Kapinska}}{2020}]{2020AAS...23632206K}
{Kapinska} A.~D.,  2020, in American Astronomical Society Meeting Abstracts
  \#236. p. 322.06

\bibitem[\protect\citeauthoryear{{Lacy} et~al.,}{{Lacy}
  et~al.}{2020}]{2020PASP..132c5001L}
{Lacy} M.,  et~al., 2020, \mn@doi [\pasp] {10.1088/1538-3873/ab63eb}, \href
  {https://ui.adsabs.harvard.edu/abs/2020PASP..132c5001L} {132, 035001}

\bibitem[\protect\citeauthoryear{{McEwen} \& {Wiaux}}{{McEwen} \&
  {Wiaux}}{2011}]{2011MNRAS.413.1318M}
{McEwen} J.~D.,  {Wiaux} Y.,  2011, \mn@doi [\mnras]
  {10.1111/j.1365-2966.2011.18217.x}, \href
  {https://ui.adsabs.harvard.edu/abs/2011MNRAS.413.1318M} {413, 1318}

\bibitem[\protect\citeauthoryear{{McMullin}, {Waters}, {Schiebel}, {Young}  \&
  {Golap}}{{McMullin} et~al.}{2007}]{2007ASPC..376..127M}
{McMullin} J.~P.,  {Waters} B.,  {Schiebel} D.,  {Young} W.,   {Golap} K.,
  2007, in {Shaw} R.~A.,  {Hill} F.,   {Bell} D.~J.,  eds,  Astronomical
  Society of the Pacific Conference Series Vol. 376, Astronomical Data Analysis
  Software and Systems XVI. p.~127

\bibitem[\protect\citeauthoryear{{Offringa} et~al.,}{{Offringa}
  et~al.}{2014}]{2014MNRAS.444..606O}
{Offringa} A.~R.,  et~al., 2014, \mn@doi [\mnras] {10.1093/mnras/stu1368},
  \href {https://ui.adsabs.harvard.edu/abs/2014MNRAS.444..606O} {444, 606}

\bibitem[\protect\citeauthoryear{{Perley}}{{Perley}}{1989}]{1989ASPC....6..259P}
{Perley} R.~A.,  1989, in {Perley} R.~A.,  {Schwab} F.~R.,   {Bridle} A.~H.,
  eds,  Astronomical Society of the Pacific Conference Series Vol. 6, Synthesis
  Imaging in Radio Astronomy. p.~259

\bibitem[\protect\citeauthoryear{{Pratley}, {Johnston-Hollitt}  \&
  {McEwen}}{{Pratley} et~al.}{2019}]{2019ApJ...874..174P}
{Pratley} L.,  {Johnston-Hollitt} M.,   {McEwen} J.~D.,  2019, \mn@doi [\apj]
  {10.3847/1538-4357/ab0a05}, \href
  {https://ui.adsabs.harvard.edu/abs/2019ApJ...874..174P} {874, 174}

\bibitem[\protect\citeauthoryear{{Pratley}, {Johnston-Hollitt}  \&
  {McEwen}}{{Pratley} et~al.}{2020}]{2020PASA...37...41P}
{Pratley} L.,  {Johnston-Hollitt} M.,   {McEwen} J.~D.,  2020, \mn@doi [\pasa]
  {10.1017/pasa.2020.28}, \href
  {https://ui.adsabs.harvard.edu/abs/2020PASA...37...41P} {37, e041}

\bibitem[\protect\citeauthoryear{{Rujopakarn} et~al.,}{{Rujopakarn}
  et~al.}{2016}]{2016ApJ...833...12R}
{Rujopakarn} W.,  et~al., 2016, \mn@doi [\apj] {10.3847/0004-637X/833/1/12},
  \href {https://ui.adsabs.harvard.edu/abs/2016ApJ...833...12R} {833, 12}

\bibitem[\protect\citeauthoryear{{Schwab} \& {Cotton}}{{Schwab} \&
  {Cotton}}{1983}]{1983AJ.....88..688S}
{Schwab} F.~R.,  {Cotton} W.~D.,  1983, \mn@doi [\aj] {10.1086/113360}, \href
  {https://ui.adsabs.harvard.edu/abs/1983AJ.....88..688S} {88, 688}

\bibitem[\protect\citeauthoryear{{Selig}, {Bell}, {Junklewitz}, {Oppermann},
  {Reinecke}, {Greiner}, {Pachajoa}  \& {En{\ss}lin}}{{Selig}
  et~al.}{2013}]{2013A&A...554A..26S}
{Selig} M.,  {Bell} M.~R.,  {Junklewitz} H.,  {Oppermann} N.,  {Reinecke} M.,
  {Greiner} M.,  {Pachajoa} C.,   {En{\ss}lin} T.~A.,  2013, \mn@doi [\aap]
  {10.1051/0004-6361/201321236}, \href
  {https://ui.adsabs.harvard.edu/abs/2013A&A...554A..26S} {554, A26}

\bibitem[\protect\citeauthoryear{{Shimwell} et~al.,}{{Shimwell}
  et~al.}{2019}]{2019A&A...622A...1S}
{Shimwell} T.~W.,  et~al., 2019, \mn@doi [\aap] {10.1051/0004-6361/201833559},
  \href {https://ui.adsabs.harvard.edu/abs/2019A&A...622A...1S} {622, A1}

\bibitem[\protect\citeauthoryear{{Steininger} et~al.,}{{Steininger}
  et~al.}{2017}]{2017arXiv170801073S}
{Steininger} T.,  et~al., 2017, arXiv e-prints, \href
  {https://ui.adsabs.harvard.edu/abs/2017arXiv170801073S} {p. arXiv:1708.01073}

\bibitem[\protect\citeauthoryear{{Tasse} et~al.,}{{Tasse}
  et~al.}{2018}]{2018A&A...611A..87T}
{Tasse} C.,  et~al., 2018, \mn@doi [\aap] {10.1051/0004-6361/201731474}, \href
  {https://ui.adsabs.harvard.edu/abs/2018A&A...611A..87T} {611, A87}

\bibitem[\protect\citeauthoryear{{Tingay}, {Tremblay}, {Walsh}  \&
  {Urquhart}}{{Tingay} et~al.}{2016}]{2016ApJ...827L..22T}
{Tingay} S.~J.,  {Tremblay} C.,  {Walsh} A.,   {Urquhart} R.,  2016, \mn@doi
  [\apjl] {10.3847/2041-8205/827/2/L22}, \href
  {https://ui.adsabs.harvard.edu/abs/2016ApJ...827L..22T} {827, L22}

\bibitem[\protect\citeauthoryear{{Wells}}{{Wells}}{1985}]{1985daa..conf..195W}
{Wells} D.~C.,  1985, in {di Gesu} V.,  {Scarsi} L.,  {Crane} P.,  {Friedman}
  J.~H.,   {Levialdi} S.,  eds, Data Analysis in Astronomy. p.~195

\bibitem[\protect\citeauthoryear{{Whittam}, {Green}, {Jarvis}  \&
  {Riley}}{{Whittam} et~al.}{2017}]{2017MNRAS.464.3357W}
{Whittam} I.~H.,  {Green} D.~A.,  {Jarvis} M.~J.,   {Riley} J.~M.,  2017,
  \mn@doi [\mnras] {10.1093/mnras/stw2638}, \href
  {https://ui.adsabs.harvard.edu/abs/2017MNRAS.464.3357W} {464, 3357}

\bibitem[\protect\citeauthoryear{{Ye}, {Gull}, {Tan}  \& {Nikolic}}{{Ye}
  et~al.}{2020}]{2020MNRAS.491.1146Y}
{Ye} H.,  {Gull} S.~F.,  {Tan} S.~M.,   {Nikolic} B.,  2020, \mn@doi [\mnras]
  {10.1093/mnras/stz2970}, \href
  {https://ui.adsabs.harvard.edu/abs/2020MNRAS.491.1146Y} {491, 1146}

\bibitem[\protect\citeauthoryear{{van Weeren} et~al.,}{{van Weeren}
  et~al.}{2017}]{2017NatAs...1E...5V}
{van Weeren} R.~J.,  et~al., 2017, \mn@doi [Nature Astronomy]
  {10.1038/s41550-016-0005}, \href
  {https://ui.adsabs.harvard.edu/abs/2017NatAs...1E...5V} {1, 0005}

\makeatother
\end{thebibliography}

%%%%%%%%%%%%%%%%%%%%%%%%%%%%%%%%%%%%%%%%%%%%%%%%%%

%%%%%%%%%%%%%%%%% APPENDICES %%%%%%%%%%%%%%%%%%%%%

\appendix

\section{34-source simulated data}
Table (\ref{tab:testcard_table}) shows the locations and fluxes of the 34 simulated point sources used in the numerical experiments. Sources are scattered around the phase centre with different fluxes across the full field of view.

\begin{table}
\centering
 \caption{Locations and fluxes of 34 simulated point sources. $X$ and $Y$ are in pixel numbers, representing the distances from the image centre $(0,0)$ to the corresponding point sources.}
 \label{tab:testcard_table}
 \resizebox{\columnwidth}{!}{%
 \begin{tabular}{l|ccc|l|ccc}
  \hline
   Index & X & Y & Flux (Jy) & Index & X & Y & Flux (Jy)\\
  \hline
  1 & 0 & 0 & 2 & 18 & 0 & 270 & 1 \\ \hline
  2 & 0 & 15 & 2 & 19 & 0 & 330 & 1 \\ \hline
  3 & -120 & 180 & 2 & 20 & 330 & 0 & 1 \\ \hline
  4 & 150 & -150 & 2 & 21 & 0 & -330 & 1 \\ \hline
  5 & 300 & 90 & 2 & 22 & -330 & 0 & 1 \\ \hline
  6 & -90 & 300 & 2 & 23 & 270 & 270 & 1 \\ \hline
  7 & 90 & -90 & 1 & 24 & 270 & -270 & 1 \\ \hline
  8 & -90 & 90 & 1& 25 & -270 & 270 & 1 \\ \hline
  9 & -90 & -90 & 1 & 26 & -270 & -270 & 1 \\ \hline
  10 & 180 & 90 & 1 & 27 & 390 & 390 & 3\\ \hline
  11 & 180 & 180 & 1 & 28 & 390 & -390 & 3 \\ \hline
  12 & 180 & -180 & 1 & 29 & -390 & -390 & 3 \\ \hline
  13 & -180 & 180 & 1 & 30 & -390 & 390 & 3 \\ \hline
  14 & -180 & -180 & 1 & 31 & 345 & 0 & 2 \\ \hline
  15 & 270 & 0 & 1 & 32 & -345 & 0 & 2\\ \hline
  16 & 0 & -270 & 1 & 33 & 0 & -345 & 2 \\ \hline
  17 & -270 & 0 & 1 & 34 & 0 & 345 & 2 \\ \hline
  \hline
 \end{tabular}
 }
\end{table}

Sources 27, 28, 29 and 30 are at the four corners of the field, and are chosen to have the largest fluxes among the 34 sources. %This arrangement is chosen to test the accuracy of wide-field imaging algorithms at positions far from the phase centre.

%%%%%%%%%%%%%%%%%%%%%%%%%%%%%%%%%%%%%%%%%%%%%%%%%%

% Don't change these lines
\bsp	% typesetting comment
\label{lastpage}
\end{document}